\documentclass{article}

\usepackage{arxiv}

\usepackage[utf8]{inputenc} 
\usepackage[T1]{fontenc}    
\usepackage{hyperref}       
\usepackage{url}            
\usepackage{booktabs}       
\usepackage{amsfonts}       
\usepackage{nicefrac}       
\usepackage{microtype}      
\usepackage{lipsum}
\usepackage{graphicx}
\usepackage{amssymb}
\usepackage{amsmath}
\usepackage{commath}
\usepackage{subcaption}
\graphicspath{ {./images/} }

\title{Steganography GAN: Cracking Steganography with Cycle Generative Adversarial Networks}

\author{
 Nibraas Khan \\
  Department of Computer Science\\
  Middle Tennessee State University\\
  Murfreesboro, TN 37132 \\
  \texttt{nak2z@mtmtail.mtsu.edu} \\
   \And
 Ruj Haan \\
  Department of Computer Science\\
  Middle Tennessee State University\\
  Murfreesboro, TN 37132 \\
  \texttt{gm3g@mtmail.mtsu.edu} \\
  \And
 George Boktor \\
  Department of Computer Science\\
  Middle Tennessee State University\\
  Murfreesboro, TN 37132 \\
  \texttt{gsb3c@mtmail.mtsu.edu} \\
  \And
 Michael McComas \\
  Department of Computer Science\\
  Middle Tennessee State University\\
  Murfreesboro, TN 37132 \\
  \texttt{mrm8m@mtmail.mtsu.edu} \\
  \And
 Ramin Daneshi \\
  Department of Computer Science\\
  Middle Tennessee State University\\
  Murfreesboro, TN 37132 \\
  \texttt{rd3s@mtmtail.mtsu.edu} \\
}

\begin{document}
\maketitle
\begin{abstract}
For as long as humans have participated in the act of communication, concealing information in those communicative mediums has manifested into an art of its own. Crytographic messages, through written language or images, are a means of concealment, usually reserved for highly sensitive or compromising information. Specifically, the field of Cryptography is the construction and analysis of protocols that prevent third parties from understanding private messages. Steganography is related to Cryptography in that the goal is to obscure information using some method or algorithm, but the most important difference is that the information and the method of concealing information within Steganography both involve images--more precisely, the embedding of one image or piece of information into another image. Ever since the creation of covert communication methods, steps have been taken to crack cryptography and steganography algorithms. The desire for this rises from both human curiosity and the need to counteract adverse uses, such as encoding harmful media in inconspicuous media (phishing attack). In this paper, we succeed in cracking the Least Significant Bit (LSB) steganography algorithm using Cycle Generative Adversarial Networks (CycleGANs) and Bayesian Optimization and compare the use of CycleGANs against Convolutional Autoencoders. The results of our experiments highlight the promising nature of CycleGANs in cracking steganography and open several possible avenues of research. 
\end{abstract}

\keywords{Cryptography \and Steganography \and Cycle Generative Adversarial Networks \and Autoencoders \and Bayesian Optimization}

\section{Introduction}
Humans have been implementing steganography one way or another since at least 440 B.C. Early Greek rulers would shave the head of a slave or prisoner of war, tattoo an image or message onto his or her scalp, wait for the prisoner's hair to regrow, obscuring the message, then send the prisoner to deliver the message \cite{siper2005rise}. Although the type of steganographic samples we'll be evaluating in this paper is a little more 21st century, the underlying principle is still there: embedding information in such a way as to be undetectable by the naked eye.

On the other side of the same coin of steganography exists cryptography: the art of writing or solving codes. Cryptography can be seen in techniques for exchanging secret keys, protocols for authenticating users, electronic auctions and elections, digital cash, and more \cite{katz_introduction_2015}.For example, if a person were to encode “Are you going to the park?” and give the encoded message to the recipient who has the decoder, they could uncover the secret message. However, if said message were to get intercepted, the third party would be unable to reverse the encoding scheme to see the real message. Steganography differs slightly from Cryptography in one key aspect: the former attempts to hide any trace of covert communication taking place while the latter can, in some cases, be identified as an encrypted message.

Steganography is especially tricky to crack because anyone evaluating an encoded piece of information would have no reasonable means to assume anything suspicious is at play. This makes steganography an important part of both the cybersecurity and malware detection communities. Additionally, for cybersecurity, many government agencies use steganography as a means to send hidden data in an image while hiding the very existence of the information itself \cite{seethalakshmi_security_2016}.

There are adverse uses of steganography, as can be expected from covert communication. For example, individuals use steganography to create malware to steal information. Malicious code can be encoded in innocuous media and fool users into installing malware. These people will often do this in the form of images or links--something that appears harmless but can do major damage to a user's system. This type of steganography is regularly seen in phishing emails that contain a piece of media which, when clicked, can release a virus into the computer and steal information. 

Due to the malicious nature of particular use cases of steganography, steps to crack such techniques to protect vulnerable users of the internet are needed. Not all users are privy to the dangerous uses of techniques such as steganography to take advantage of them. An important step needed to be taken to protect internet users is cracking steganography algorithms to combat attacks such as phishing. 

The primary characteristic of our specific steganography algorithm that we crack in this paper involves taking two images, where one image is the Hidden Image and another image is the Cover Image, and encoding the Hidden Image into the Cover Image by taking the first $N$ number of significant digits in each of its binary string representations, the red-green-blue value per pixel, and appending it to the corresponding strings in the Cover Image. This particular steganography technique is known as Least Significant Bit (LSB).

Our algorithm to crack the LSB algorithm implements Cycle Generative Adversarial Networks (CycleGANs). CycleGANs are a method for deep learning to synthesize data with the use of two Generate Adversarial Networks (GANs). This strategy of image to image translation between two different domains was originally introduced in a paper by Ian Goodfellow with his fellow researches at the University of Montreal in 2014 \cite{welander_generative_2018}. In our case, the two domains will be the encoded images and the decoded images.

Our implementation of the CycleGANs will use a system called Pix2pix as the pre-trained model. Pix2pix is a non-parametric texture model for translating an image to another image. This allows our algorithm to start translation between images and start generating our desired images without having to learn everything from scratch \cite{meng_steganography_2019}.

On top of our algorithm, we implement Bayesian Optimization to fine-tune the hyper-parameters of our algorithm. This methodology is used to optimize functions with expensive evaluations.

There are numerous harmful uses of steganography used to take advantage of naive internet users, so certain steps need to be taken. We provide a technique to crack such malicious steganography. This could prove useful in fields directly related to steganography, such as cybersecurity, to other more benign fields such as photography. There is a clear need for tools to crack steganography algorithms. 

\section{Background}

To begin understanding how we approached the problem of cracking the Steganography algorithm, it would be useful to begin with Generative Adversarial Networks (GANs), generative models that are able to produce high quality content (high quality in this case meaning any kind of data that closely resembles an optimal distribution specified for a given task) through adversarial learning. GANs were first introduced in 2014 by Ian J. Goodfellow and his colleagues as a way to circumvent the common pitfalls of deep generative models at the time. They introduced 'adversarial nets' that pit two networks against each other in an attempt to refine the generative capabilities of a generator network \cite{goodfellow_generative_2014}. Specifically, GANs partake in a discrimination task wherein a discriminator network is trained to classify content as real or fake and the generator network is trained to produce content real enough to fool the discriminator network. For GANs, after each training cycle, a classification error is evaluated based on how often the discriminator failed in its discrimination task, and the goal of the generator network is to maximize this classification error while the discriminator network attempts to minimize that error. Weights are adjusted according to each network's task: the generator via gradient ascent over its parameters and the discriminator via gradient descent over its parameters. \cite{radford_unsupervised_2016, paloniemi_introduction_2020} As both networks become more proficient in their respective tasks, the adversarial nature of the training regime will allow the generator to produce more accurate content with respect to what the task demands.  

CycleGANs demonstrate a natural progression of GANs architecture. These models are useful for image-to-image translation, commonly implemented as style transfer, wherein the architecture learns to translate one image from one domain to the style or pattern of another image from another domain. Unlike regular GANs, which implement one discriminator network and one generator network to achieve realism in one given image domain, CycleGANs implement two of these networks to achieve a translation of one image domain to another. Specifically, this is achieved by training two generators and two discriminators, two distinct GAN implementations, where Model A's generator is discriminated by Model B's discriminator, and both networks are configured to their respective image domains \cite{chu_cyclegan_2017}.

Our CycleGANs architecture implements Pix2pix, a service which learns the mapping of an input image and generates a corresponding output image. Pix2pix was developed by a Dutch broadcasting network, NPO, in an attempt to create an AI system that could analyze user-created images and convert them to their life-like representation. Ideally, the model aims to translate one possible representation of a scene or image into another, but the native approach with Convolutional Neural Networks isn't ideal because they train to minimize the loss function, and therefore, produce blurry images. Generative Adversarial Networks circumvent this problem because they aim to classify fake (blurry or imprecise) information while also reducing the loss on the generator's part. The loss in a GANs architecture adapts to the data, and therefore, many different kinds of images can be generated with Pix2pix. An example of pix2pix pattern translation is taking a series of abstract squares on a canvas and transforming them into windows on a brick building. Unlike pure CycleGANs, which, for our implementation, is useful for translating styles or patterns from one concrete image to another, Pix2pix aims to translate abstract patterns into concrete images, and can emulate that process in reverse, and can even do things like remove backgrounds from images and convert edge drawings into life-like pictures of cats \cite{isola_image--image_2018}!

Pix2pix serves as a way to generate images with the use of an ordered set meaning each abstract image has a one-to-one correspondence with a concrete image. CycleGANs attempt to solve the problem of un-ordered sets meaning there is no one-to-one pairing of images. We can use the models in Pix2pix as pre-trained models to use in our CycleGAN algorithm.

Current literature is also attempting to crack cryptography, one example of this is Cipher GANs. Related to Cycle GANs, Cipher GANs, a type of model that can understand the underlying cipher mapping of unordered data given the encrypted and unecrypted corpus, are more useful in discrete tasks such as text alignment because they more effectively address the problem of flat gradients through "discrete variables" and they also solve the problem of uninformative discrimination in a GANs architecture wherein the discriminator favors a discrimination criterion that is uninformative when compared to the re-discretized samples from the generator. The philosophy of Cipher GANs is particularly important to us because our problem domains are similar. Unlike Cipher GANs, which aim to crack discretized data in the form of text and cipher sequences, we aim to crack embedded images by learning the steganographic alogrithm with which they're encoded \cite{gomez_unsupervised_2018}.

To show the effectiveness of Cycle GANs, we implement another machine learning algorithm to crack steganography. Autoencoders are a type of unsupervised learning model that is particularly good at feature extraction by trying to approximate an identity function that produces target values similar to the input values of the network \cite{ng2011sparse}. Autoencoders, along with other unsupervised learning techniques, were discussed as early as 1987 as a proposed method for unsupervised pre-training of artificial neural networks \cite{schmidhuber_deep_2015}. We're attempting to use Convolutional Autoencoders to demonstrate its similar utility to Cycle GANs for cracking Steganography due to this type of model's ability to learn embedded features in a dataset \cite{guo_deep_2017}.

To further improve the accuracy of our model we implement Bayesian Optimization which uses Bayesian inference to improve the hyper-parameters. Bayesian optimization, coined by Jonas Mockus from the 1970s-1980s, is a sequential design strategy for global optimization of a black-box function that doesn't require derivatives \cite{mockus_bayes_1974, mockus_bayes_1975}. This is particularly important for hyper-parameter tuning as there is no way to take a derived of the model. Bayesian Optimization assumes that the noise in the problem is in the hyper-parameters, so the algorithm attempts to find the best combination of values to maximize model output accuracy.

\section{Methods}

There are several components to CycleGANs and our implementation of it. There is a certain degree of flexibility with our components, but we will present the implementation of our steganography algorithm, CycleGAN, Autoencoder, and our testing protocols. 

\subsection{Steganography algorithm}
There have been many Steganography techniques created to encode information in images \cite{johnson_exploring_1998}. Some of them are complex and difficult to crack, while others are done naively and allow for easy cracking. 

\begin{figure}[ht]
    \centering
    \includegraphics[scale=0.13,clip=false]{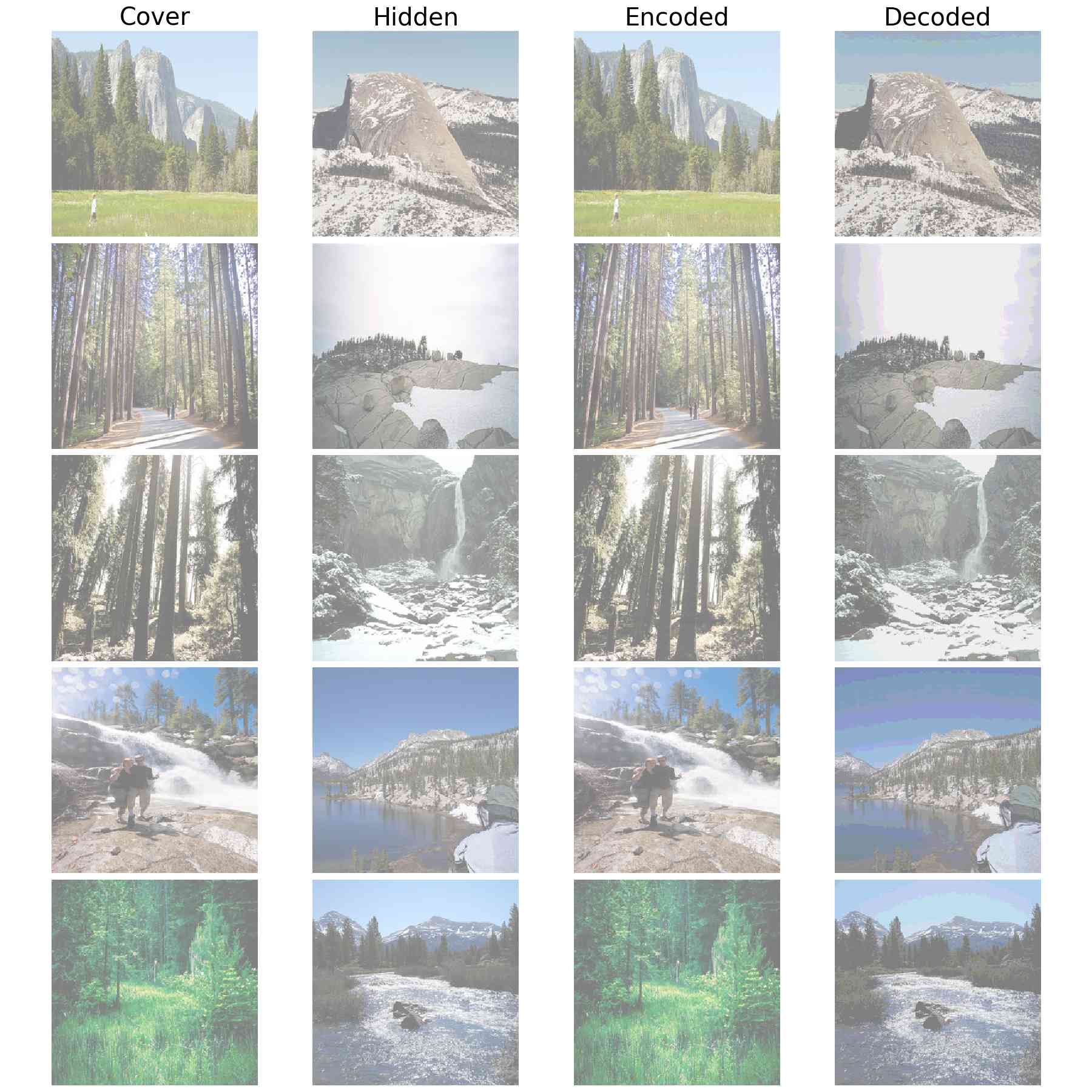}
    \caption{Implementation of the LSB algorithm where the encoded image shows the cover image with the hidden image inside it, and the decoded image shows the extracted hidden image from the encoded image.}
    \label{fig:LSB}
\end{figure}

One of the most common Steganography techniques is Least Significant Bit (LSB), a technique where the hidden information is hidden in the least significant bits of the cover file \cite{singh2015steganography}. This is achieved by replacing $X$ number of the least significant bits of the cover file and replacing it with $8-X$ of most significant bits from the hidden image for each RGB value. The results of the LSB method is shown in Figure \ref{fig:LSB}.

\subsection{Cycle Generative Adversarial Network Model Description}
Traditionally, image to image translation has required paired training examples such as a direct translation from an image of a horse to an image of a zebra. A well-known implementation of paired image to image translation has been Pix2pix \cite{isola_image--image_2018}. Often times, these kinds of datasets are not available or cannot be created. There is no way to create an ordered set of images of famous paintings and their respective photographs. 

CycleGANs are used for image to image translation for unordered sets; they take an image from one domain and convert it to another domain such as horses to zebras. There are other techniques for image to image translation, but CycleGANs allow for unsupervised training without paired examples. 

This model learns to translate an image from a source domain $X$ to a target domain $Y$ without paired examples. The mapping $G: X$ $\rightarrow$ $Y$ is learned such that the distribution of images from $G(X)$ is indistinguishable from the distribution $Y$. Furthermore, there is an inverse mapping function $F: Y$ $\rightarrow$ $X$ and a cycle consistency loss to push $F(G(X))$ $\sim$ $X$ and $G(F(Y))$ $\sim$ $Y$ \cite{zhu_unpaired_2018}.

The GAN architecture of one generator model and one discriminator model is extended to simultaneously train two generator models, $G$ and $F$, and two discriminator models, $D_Y$ and $D_X$. Generator $G$ takes an image from domain $X$ and maps it to domain $Y$, generator $F$ takes an image from domain $Y$ and maps it to domain $X$, and the two discriminators check the plausibility of the generated images with their respective generator. Furthermore, there is a cycle consistency loss which dictates that the output of $G$ can be used as the input of $F$ and the output of $F$ should match the original input. 

The objective function for the mapping function $G: X$ $\rightarrow$ $Y$ is expressed as: 

\begin{align}
    \begin{split}
        \mathcal{L}_{GAN}(G,D_Y,X,Y) &= \mathbb{E}_{y \sim p_{data}(y)} [\log D_Y(y)]  \\
        &+ \mathbb{E}_{x \sim p_{data}(x)} [\log(1-D_Y(G(x)))]
    \end{split}
\end{align}

where $G$ tries to generate images that look similar to the images from domain $Y$, and $D_Y$ tries to distinguish between the generated images and the real samples. This is essentially a Minimax game as $G$ aims to minimize the objective function and the adversary, $D$, tries to maximize it. 

Similarly, there is an adversarial loss for the mapping function $F$:

\begin{align}
    \begin{split}
        \mathcal{L}_{GAN}(G,D_X,Y,X) &= \mathbb{E}_{x \sim p_{data}(x)} [\log D_X(x)]  \\
        &+ \mathbb{E}_{y \sim p_{data}(y)} [\log(1-D_X(G(y)))]
    \end{split}
\end{align}

where $F$ tries to generate images that look similar to the images from domain $X$, and $D_X$ tries to distinguish between the generated images and the real samples.

\begin{figure*}[!hbt]
\centering
\textbf{Autoencoder}\par\medskip
    \begin{subfigure}[b]{0.28\textwidth}
    \centering
            \includegraphics[scale=0.13,clip=false]{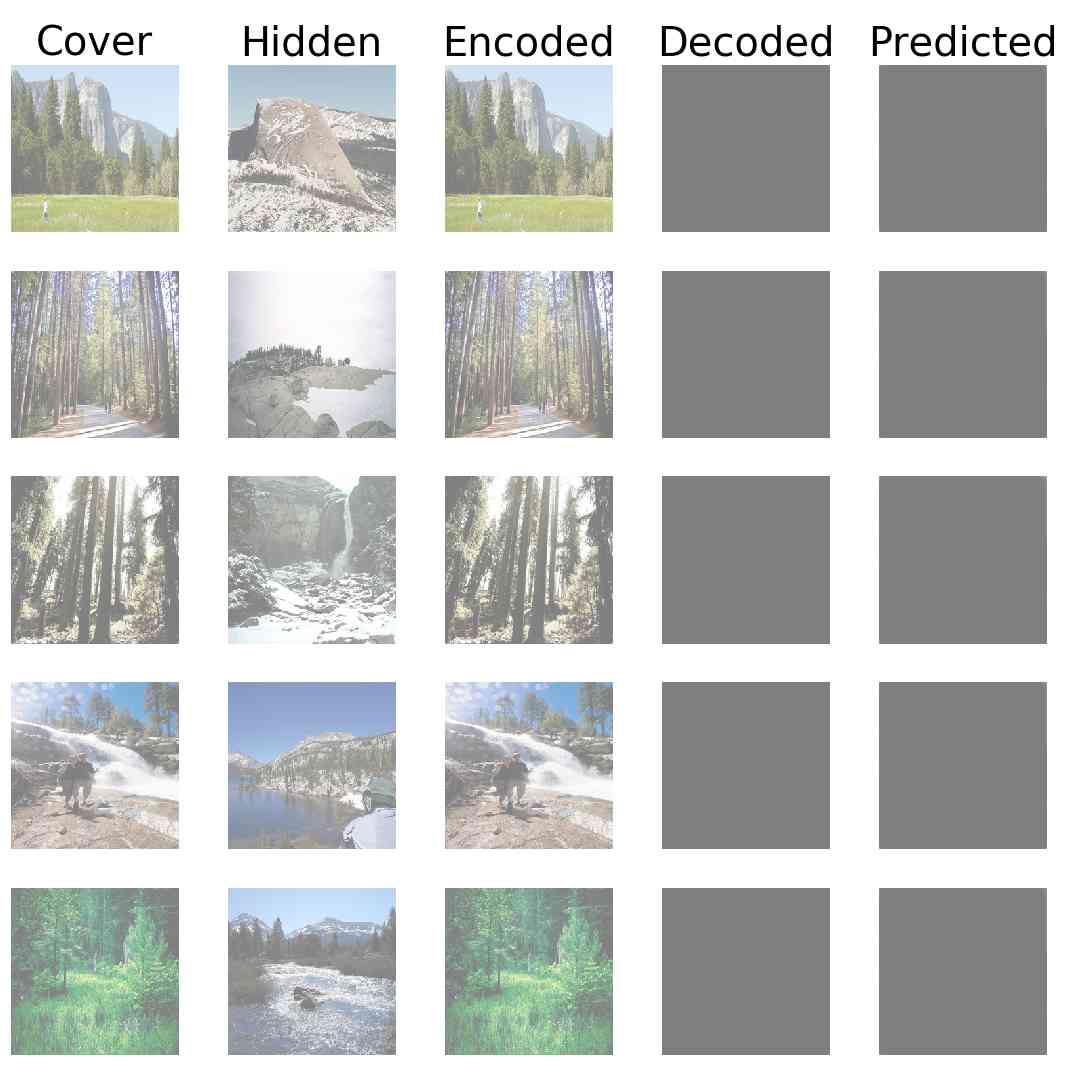}
            \caption{Autoencoder trained on 0 bits}
            \label{autoencoder_0}
    \end{subfigure}
    \hspace{0.05\textwidth}
    \begin{subfigure}[b]{0.28\textwidth}
    \centering
            \includegraphics[scale=0.13,clip=false]{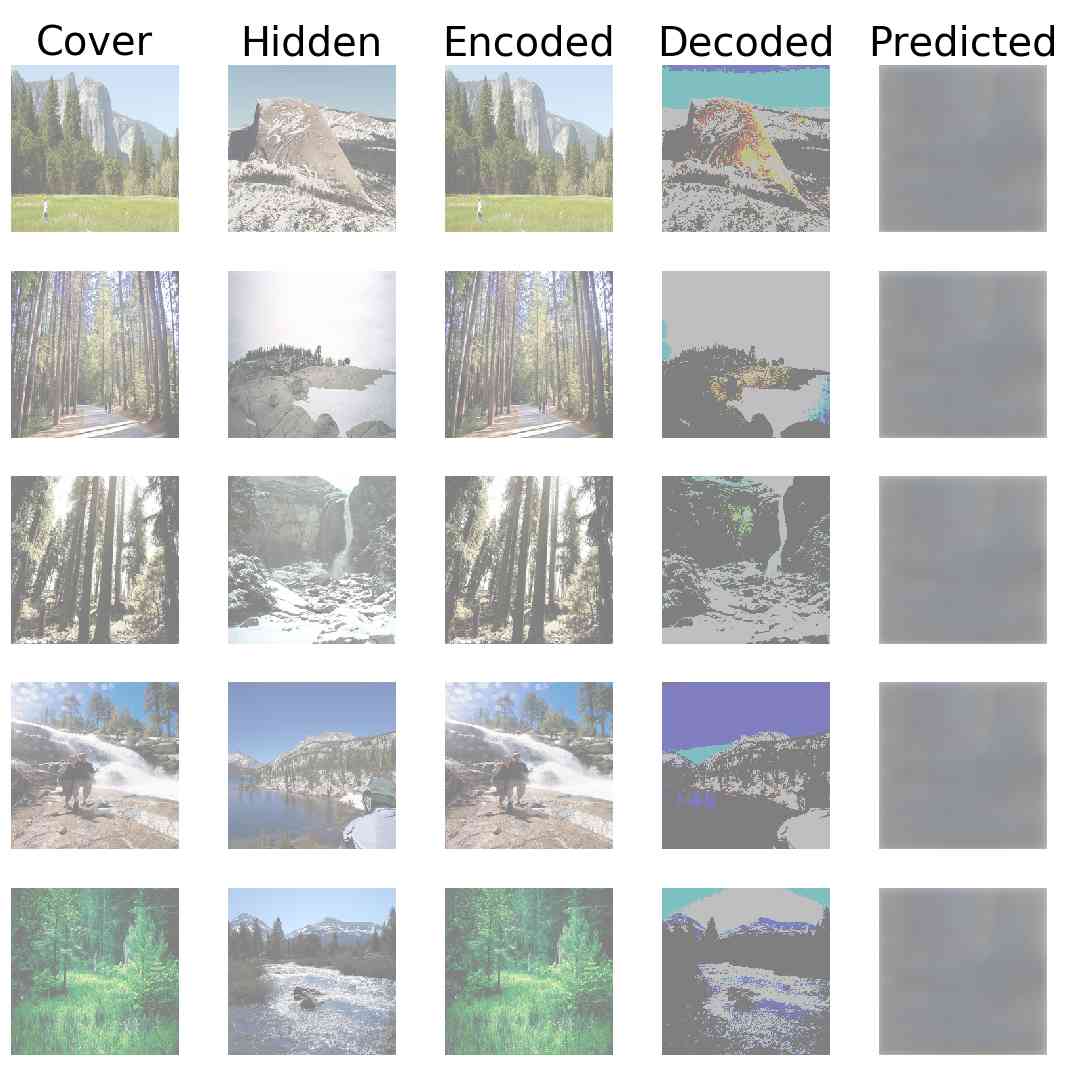}
            \caption{Autoencoder trained on 1 bit}
            \label{autoencoder_1}
    \end{subfigure}
    \hspace{0.05\textwidth}
    \begin{subfigure}[b]{0.28\textwidth}
    \centering
            \includegraphics[scale=0.13,clip=false]{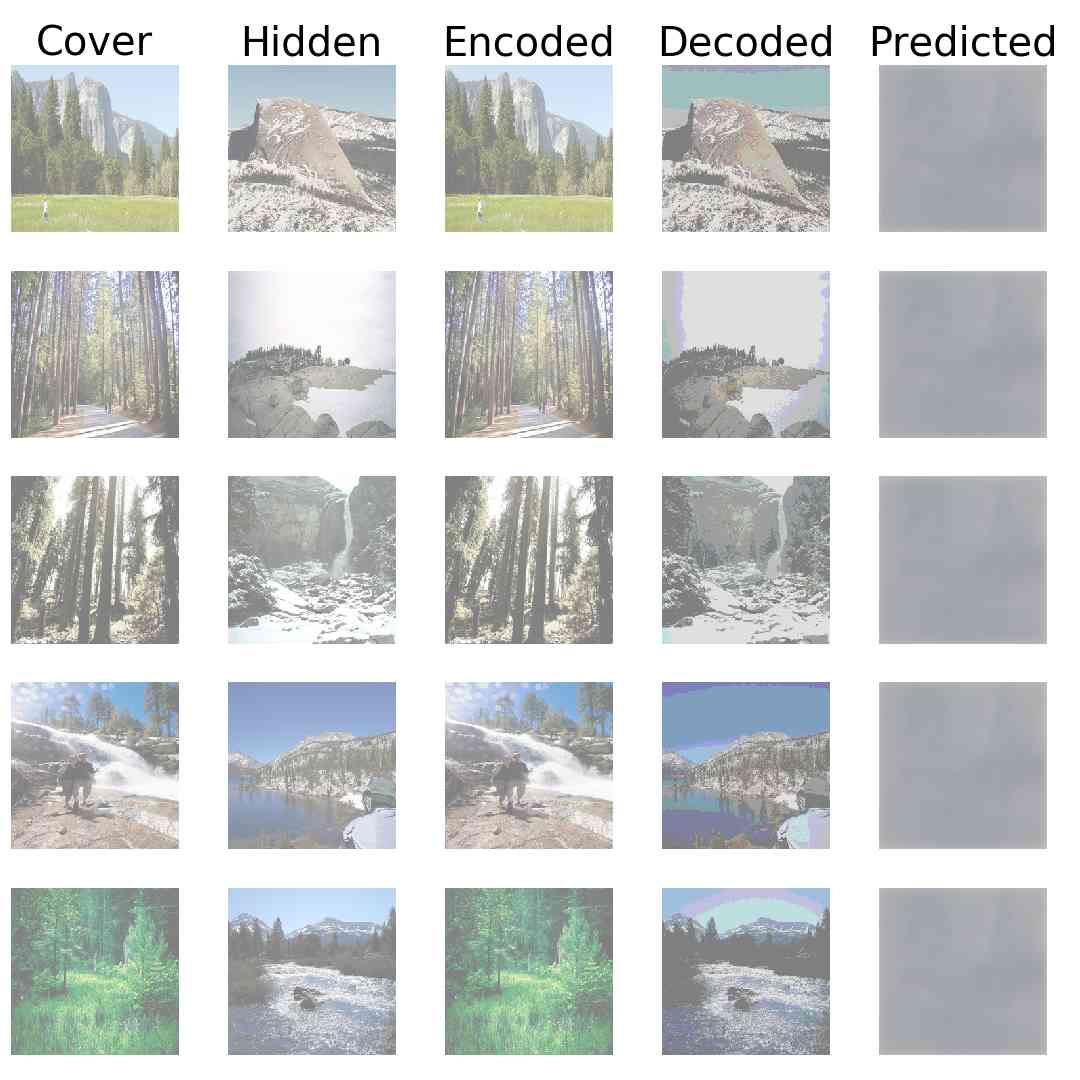}
            \caption{Autoencoder trained on 2 bits}
            \label{autoencoder_2}
    \end{subfigure}
    \hspace{0.05\textwidth}
    \begin{subfigure}[b]{0.28\textwidth}
    \centering
            \includegraphics[scale=0.13,clip=false]{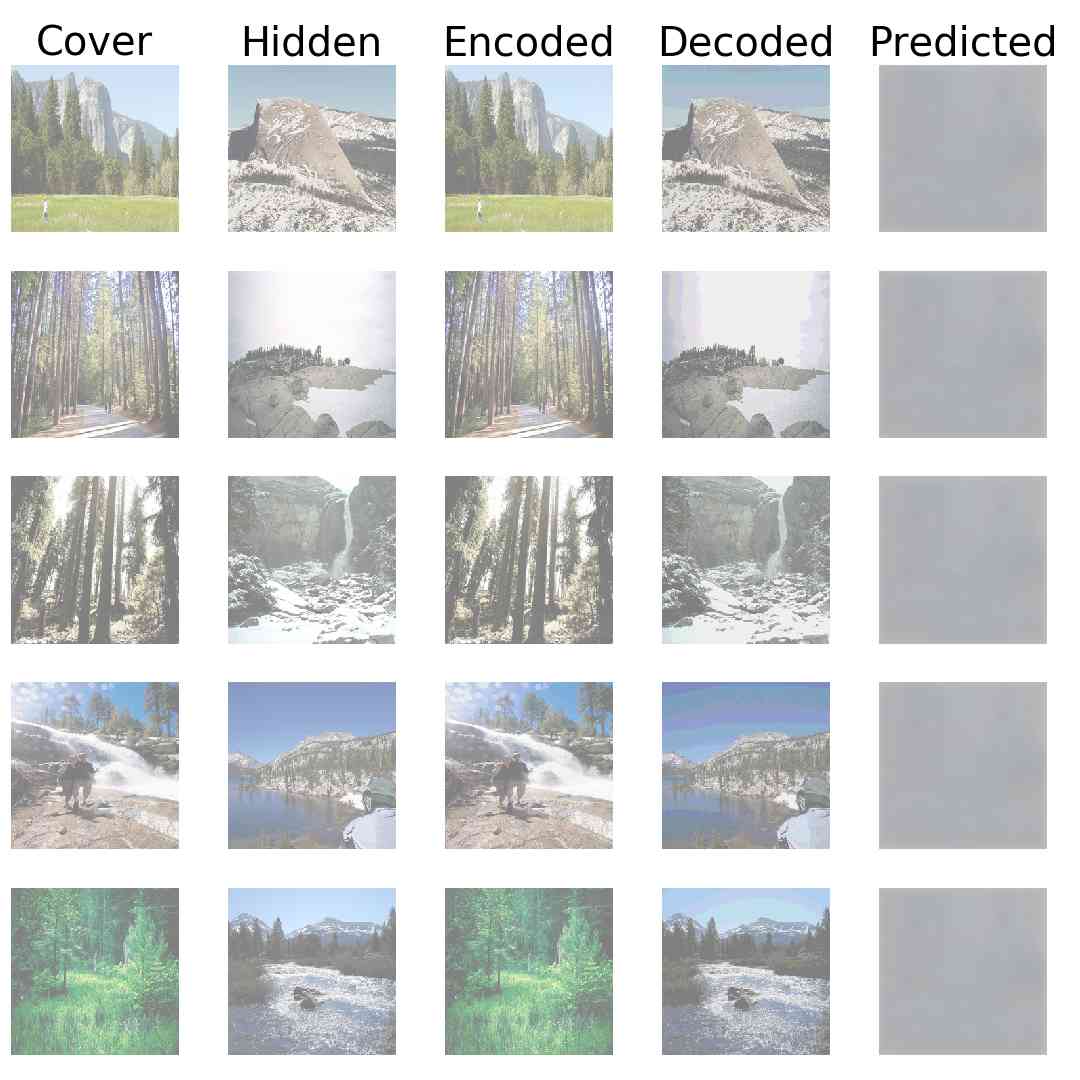}
            \caption{Autoencoder trained on 3 bits}
            \label{autoencoder_3}
    \end{subfigure}
    \hspace{0.05\textwidth}
    \begin{subfigure}[b]{0.28\textwidth}
    \centering
            \includegraphics[scale=0.13,clip=false]{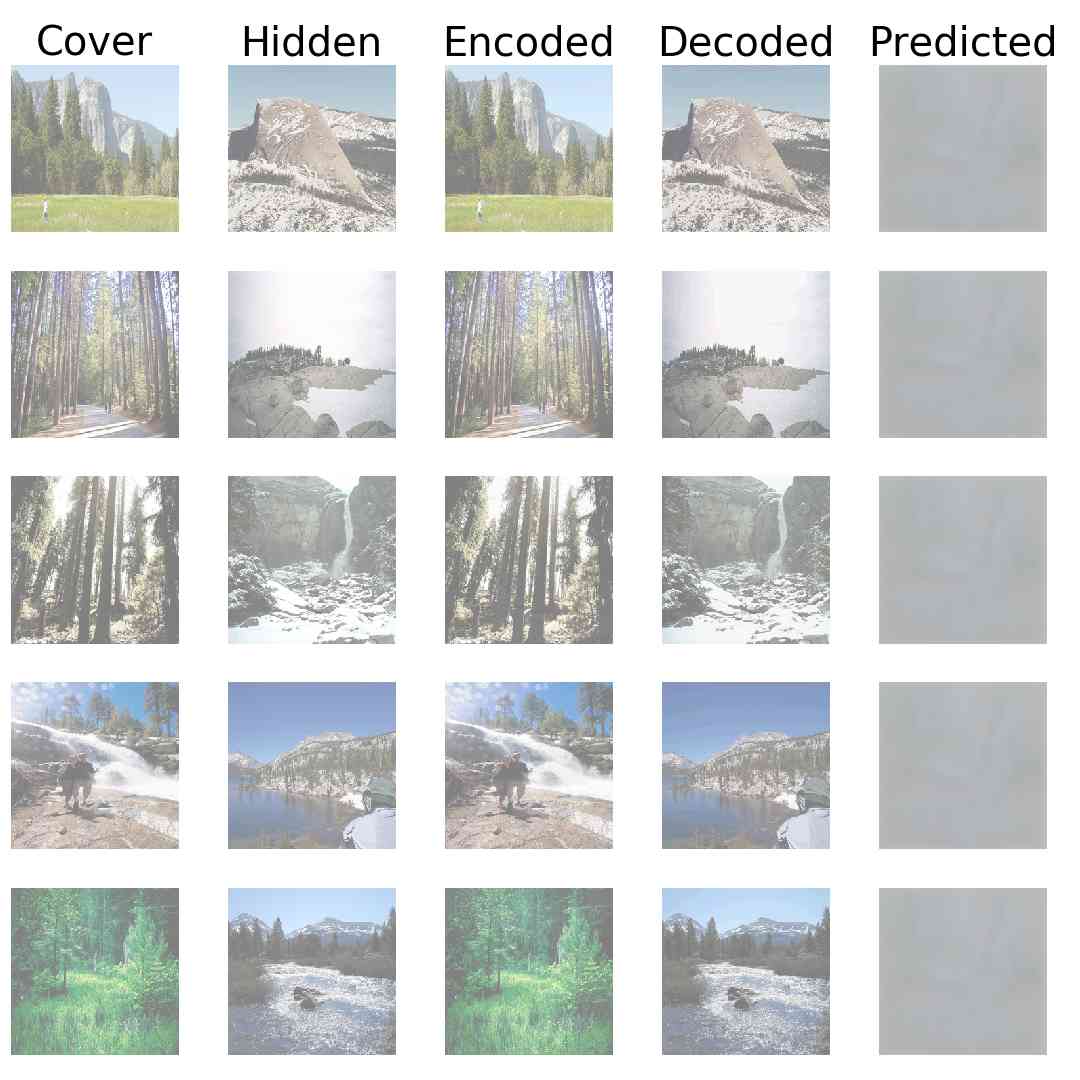}
            \caption{Autoencoder trained on 4 bits}
            \label{autoencoder_4}
    \end{subfigure}
    \hspace{0.05\textwidth}
    \begin{subfigure}[b]{0.28\textwidth}
    \centering
            \includegraphics[scale=0.13,clip=false]{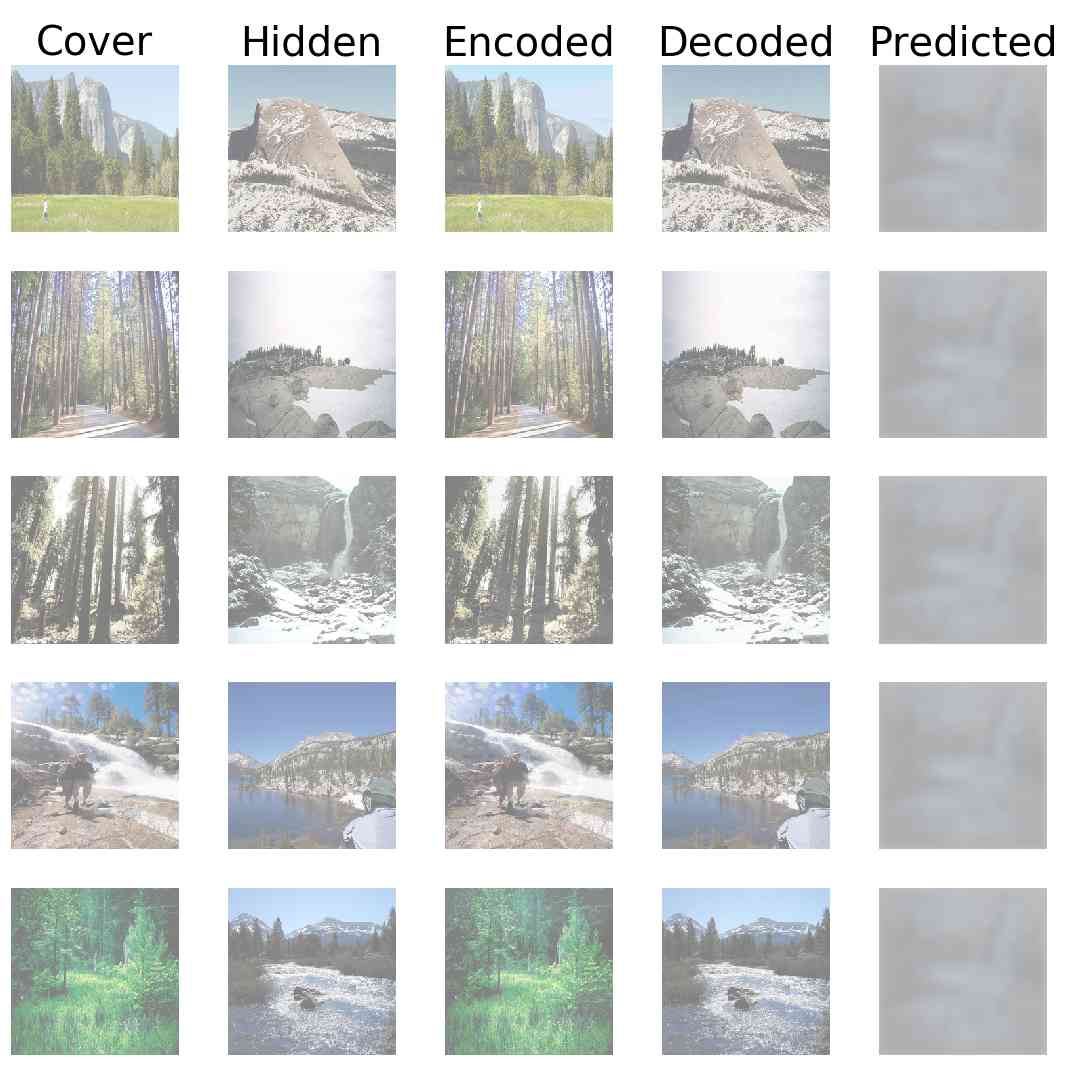}
            \caption{Autoencoder trained on 5 bits}
            \label{autoencoder_5}
    \end{subfigure}
    \hspace{0.05\textwidth}
    \begin{subfigure}[b]{0.28\textwidth}
    \centering
            \includegraphics[ scale=0.13,clip=false]{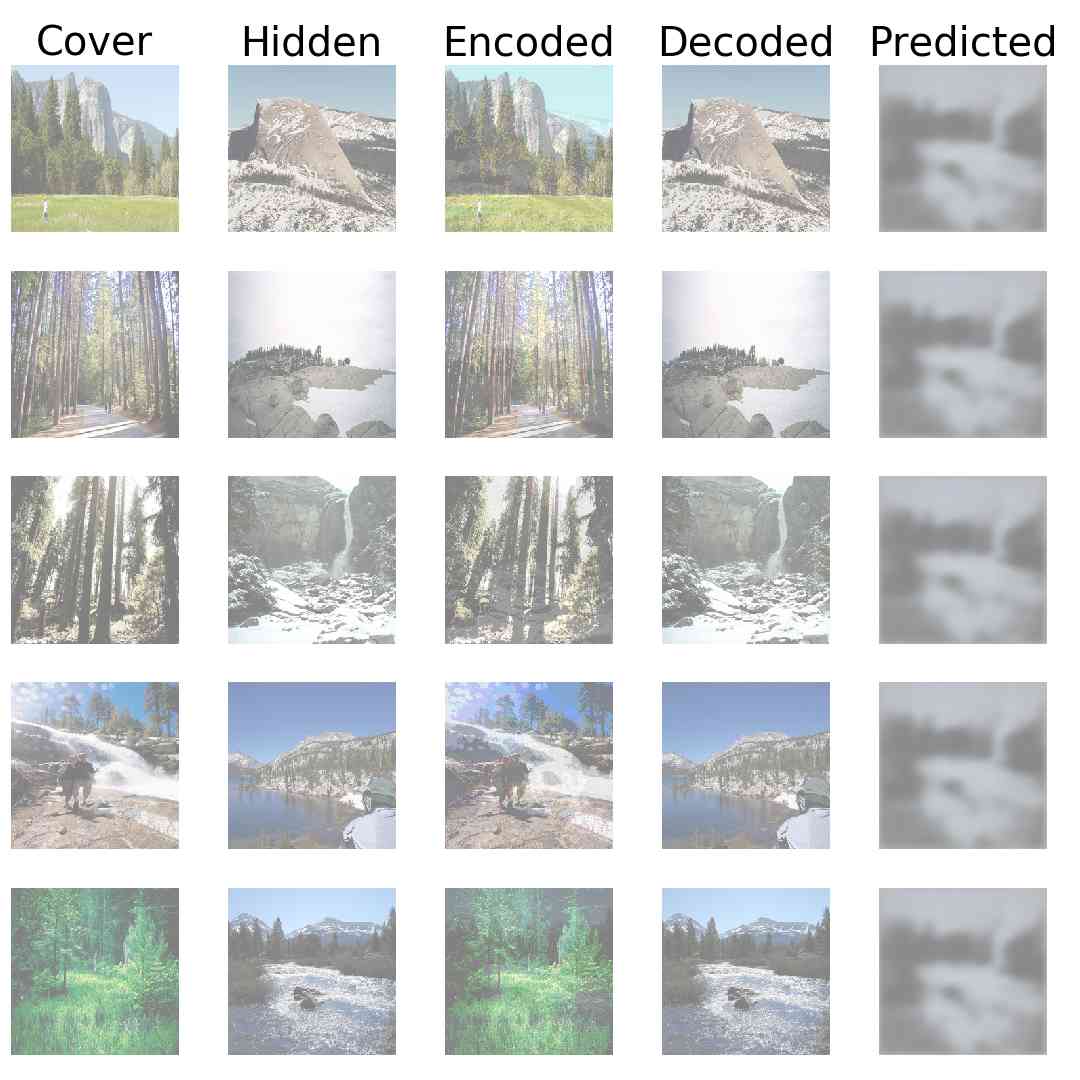}
            \caption {Autoencoder trained on 6 bits}
            \label{autoencoder_6}
    \end{subfigure}
    \hspace{0.05\textwidth}
    \begin{subfigure}[b]{0.28\textwidth}
    \centering
            \includegraphics[scale=0.13,clip=false]{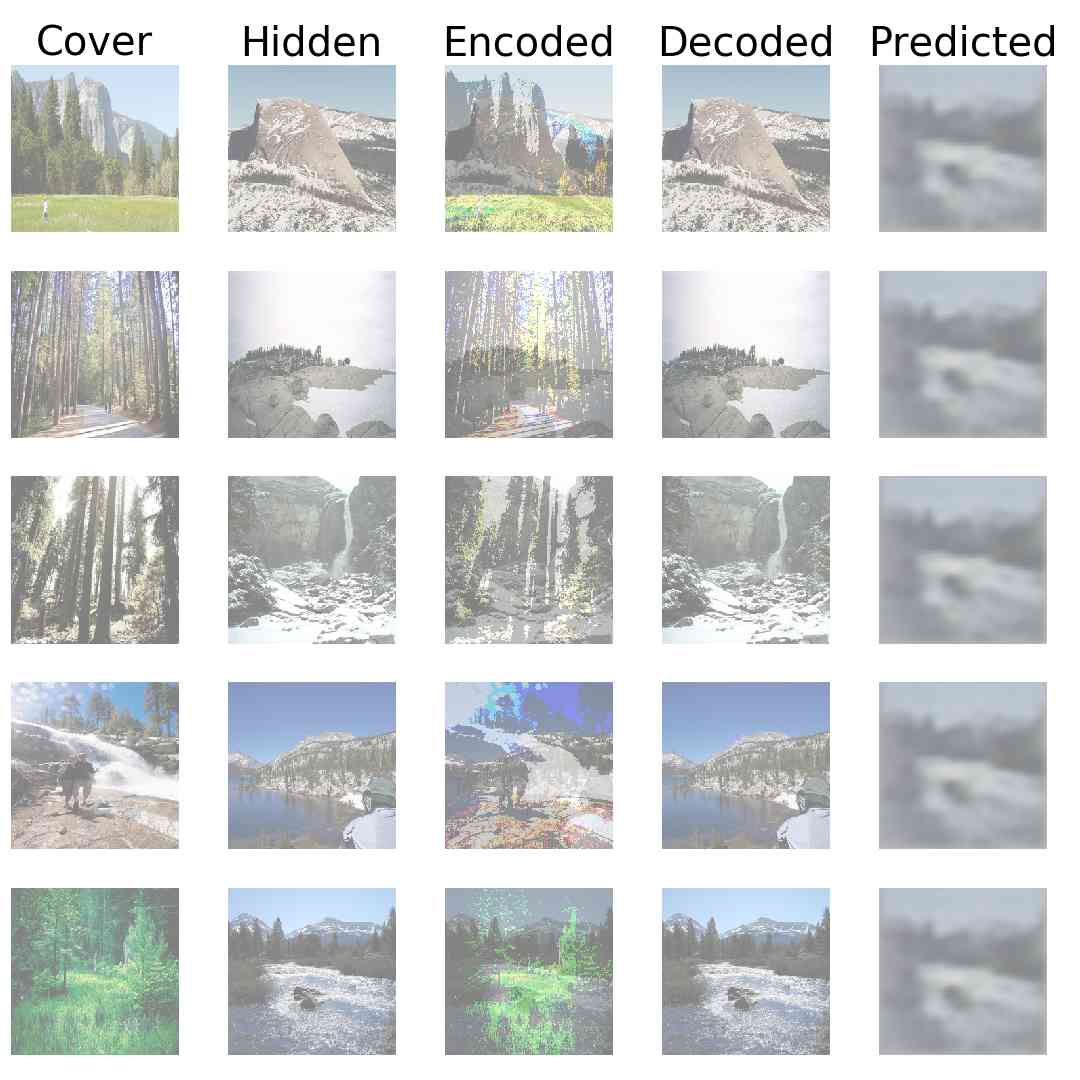}
            \caption {Autoencoder trained on 7 bits}
            \label{autoencoder_7}
    \end{subfigure}
    \hspace{0.05\textwidth}
    \begin{subfigure}[b]{0.28\textwidth}
    \centering
            \includegraphics[ scale=0.13,clip=false]{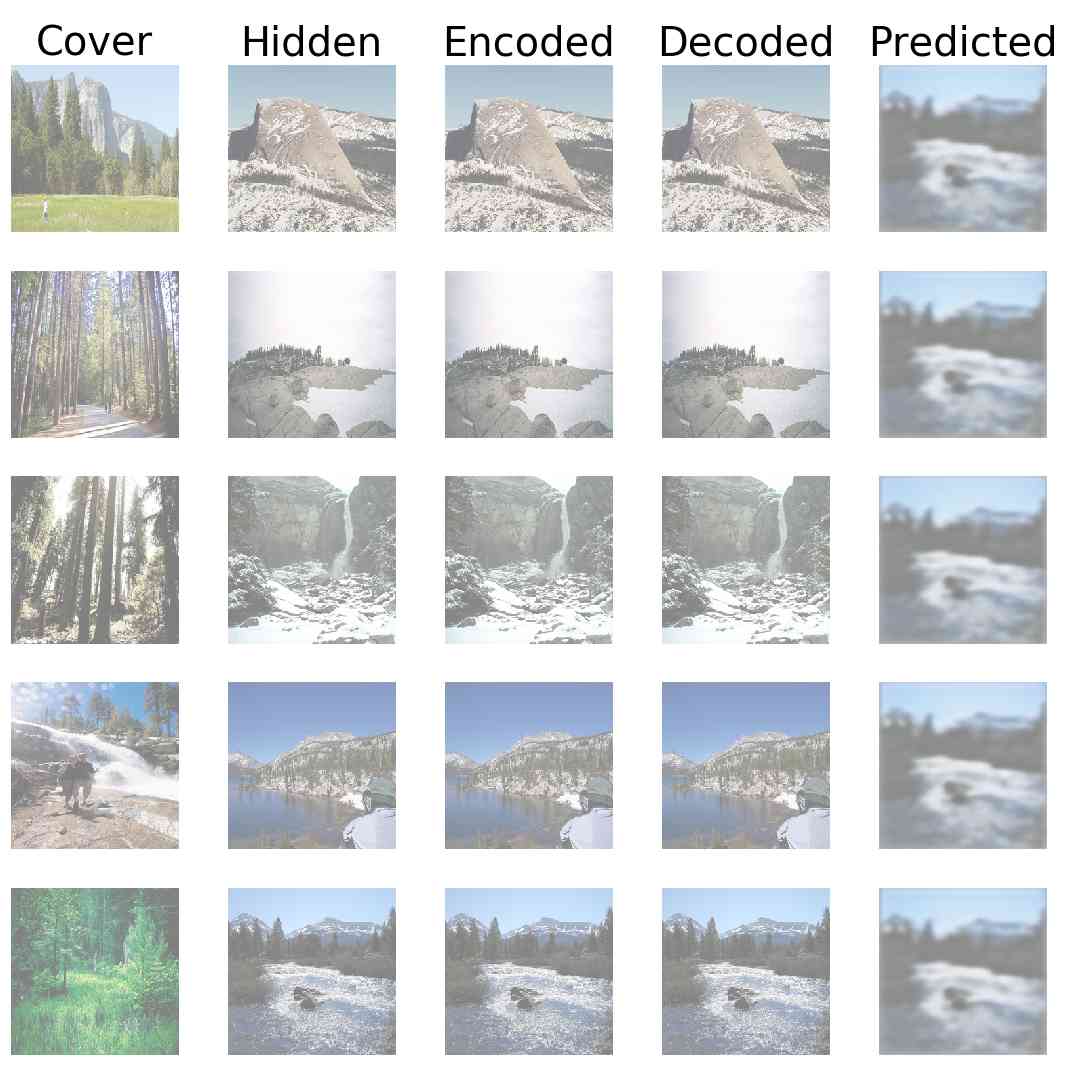}
            \caption {Autoencoder trained on 8 bits}
            \label{autoencoder_8}
    \end{subfigure}
\caption{Each of the images corresponds to an independent Autoencoder model trained on a specific number of hidden bit sizes. Each individual column of the images corresponds to the cover images from domain one, hidden images from domain two, encoded images using that specific bit size, decoded images using that specific bit size, and the predicted decoded image by the Autoencoder, respectively. The Autoencoders try to predict five images from its training set.}
\label{fig:autoencoder_images}
\end{figure*}

Adversarial training can learn to solve the problem of image to image translation; however, with a large enough search space, a network can take an input image and map it to any number of permutations from the target domain since the distribution matches. So there is no guarantee that the model will be able to learn to appropriately match the input to the output. To limit the search space, cycle consistency loss is introduced. 

\begin{align}
    \begin{split}
        \mathcal{L}_{cyc}(G, F) &= \mathbb{E}_{x \sim p_{data}(x)}[{\lVert F(G(x)) - x \rVert}_1] \\
        &+ \mathbb{E}_{y \sim p_{data}(y)}[{\lVert G(F(y)) - y \rVert}_1]
    \end{split}
\end{align}

where $x$ $\rightarrow$ $G(x)$ $\rightarrow$ $F(G(x))$ $\approx$ $x$ and $y$ $\rightarrow$ $F(y)$ $\rightarrow$ $G(F(y))$ $\approx$ $y$.

The full objective can be written as:

\begin{align}
    \begin{split}
        \mathcal{L}(G,F,D_X,D_Y) &= \mathcal{L}_{GAN}(G,D_Y,X,Y) \\
        &+ \mathcal{L}_{GAN}(F,D_X,Y,X) \\
        &+ \lambda \mathcal{L}_{cyc}(G,F)
    \end{split}
\end{align}

where $\lambda$ controls the importance of the objectives. The model plays a large Minimax game and tries to solve:

\begin{align}
    \begin{split}
        G^*, F^* = {\arg}\:\underset{G, F}{\min}\:\underset{D_x, D_Y}{\max}\:\mathcal{L}(G,F,D_X,D_Y)
    \end{split}
\end{align}

By solving the equation above, the model effectively learns to translate between domain $X$ and domain $Y$ using two generators and two discriminators with the use of cycle consistency loss. 

The CycleGAN algorithm solves its objective, and learns how to translate between the encoded steganography domain and the decoded steganography domain. 

\begin{figure*}[!hbt]
\centering
\textbf{CycleGAN Training}\par\medskip
    \begin{subfigure}[b]{0.28\textwidth}
    \centering
            \includegraphics[scale=0.08]{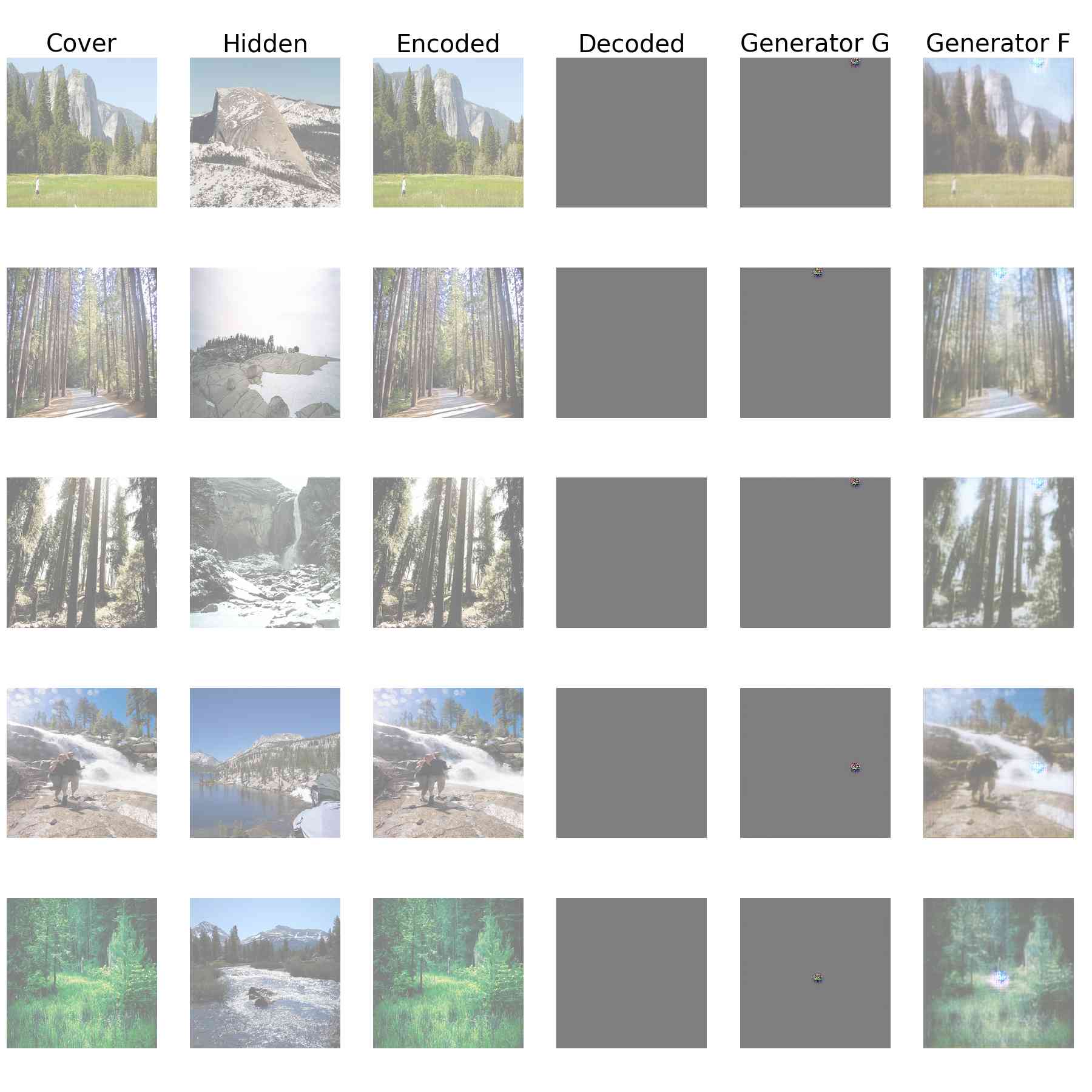}
            \caption{CycleGAN trained on 0 bits}
            \label{cycle_gan_0}
    \end{subfigure}
    \hspace{0.05\textwidth}
    \begin{subfigure}[b]{0.28\textwidth}
    \centering
            \includegraphics[scale=0.08]{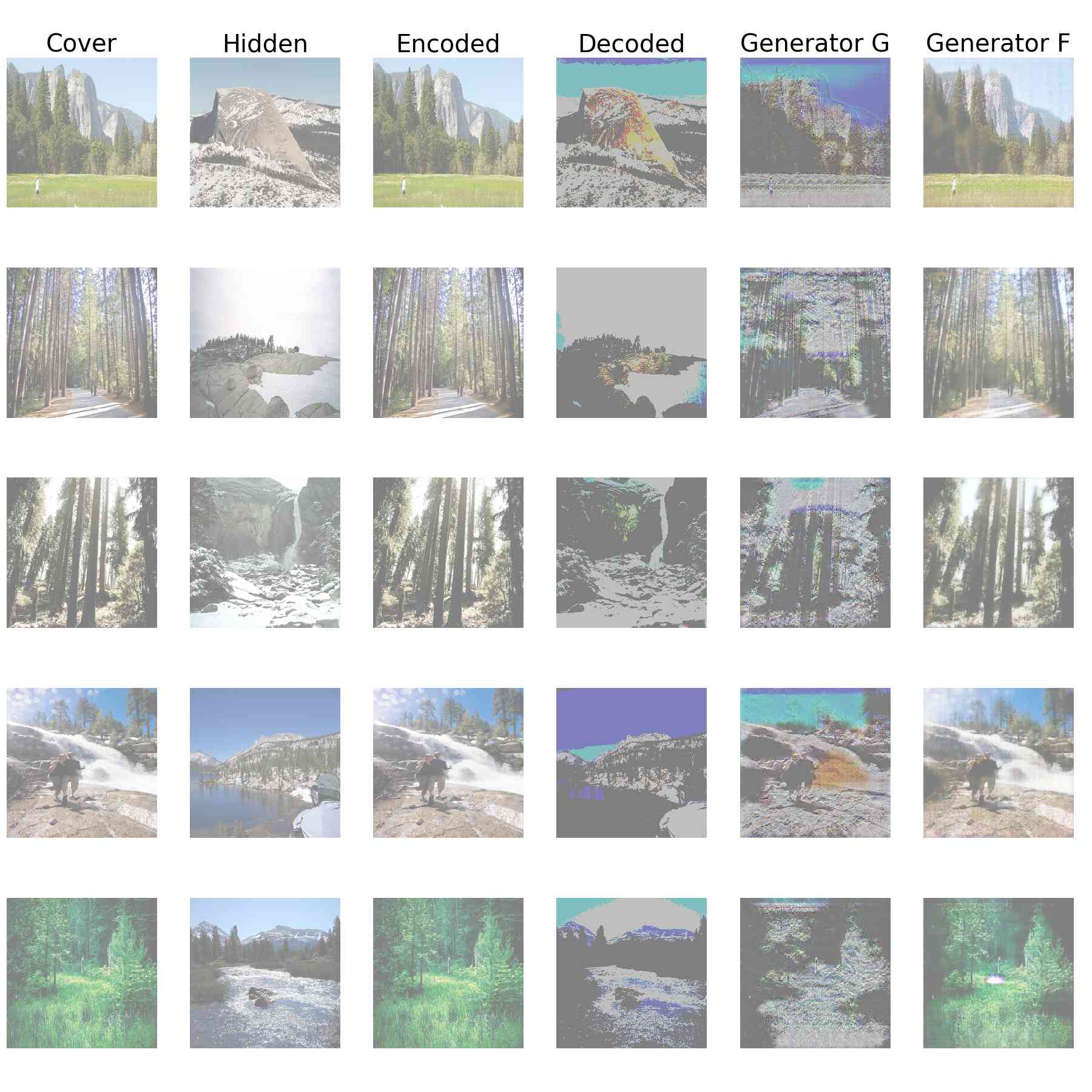}
            \caption{CycleGAN trained on 1 bit}
            \label{cycle_gan_1}
    \end{subfigure}
    \hspace{0.05\textwidth}
    \begin{subfigure}[b]{0.28\textwidth}
    \centering
            \includegraphics[scale=0.08]{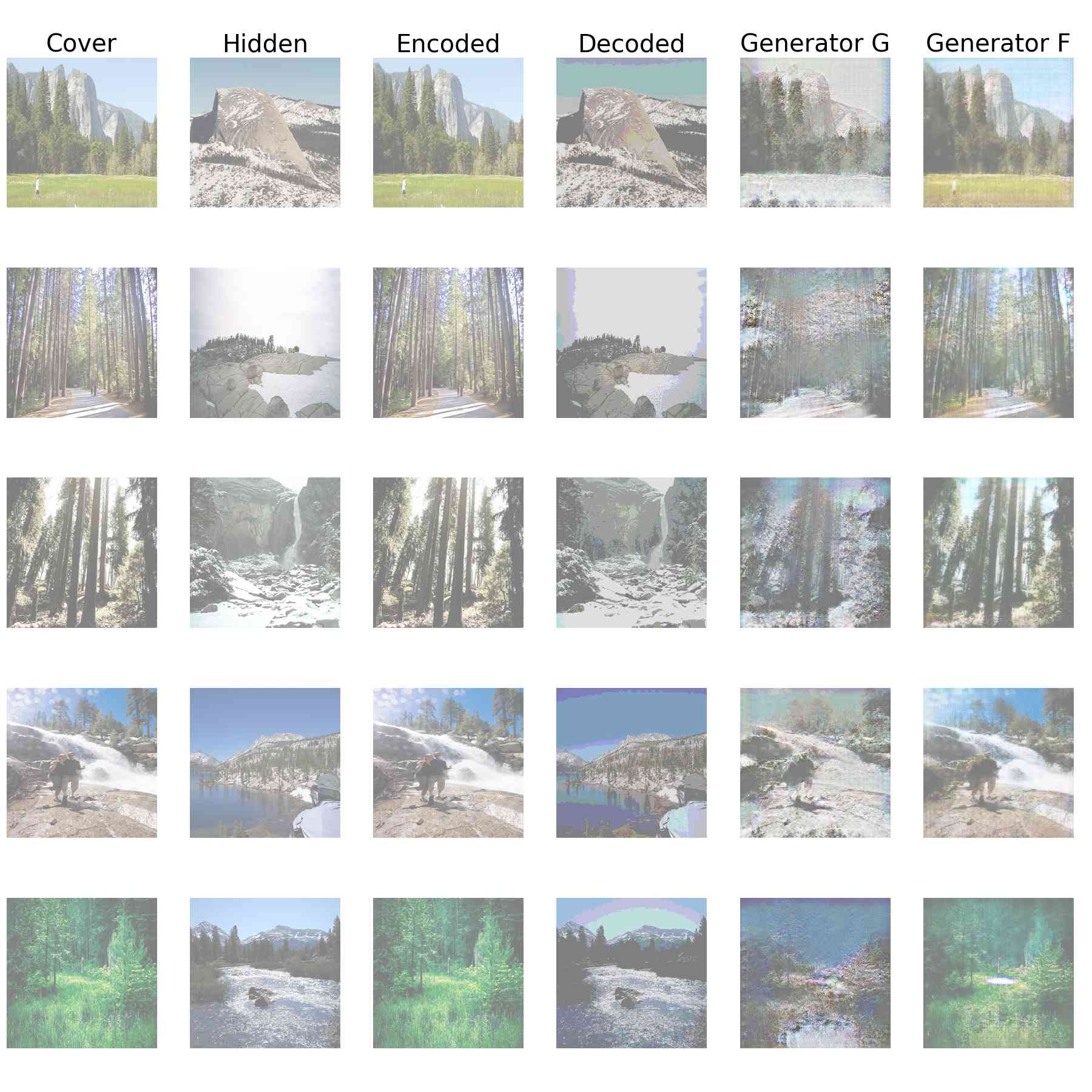}
            \caption{CycleGAN trained on 2 bits}
            \label{cycle_gan_2}
    \end{subfigure}
    \hspace{0.05\textwidth}
    \begin{subfigure}[b]{0.28\textwidth}
    \centering
            \includegraphics[scale=0.08]{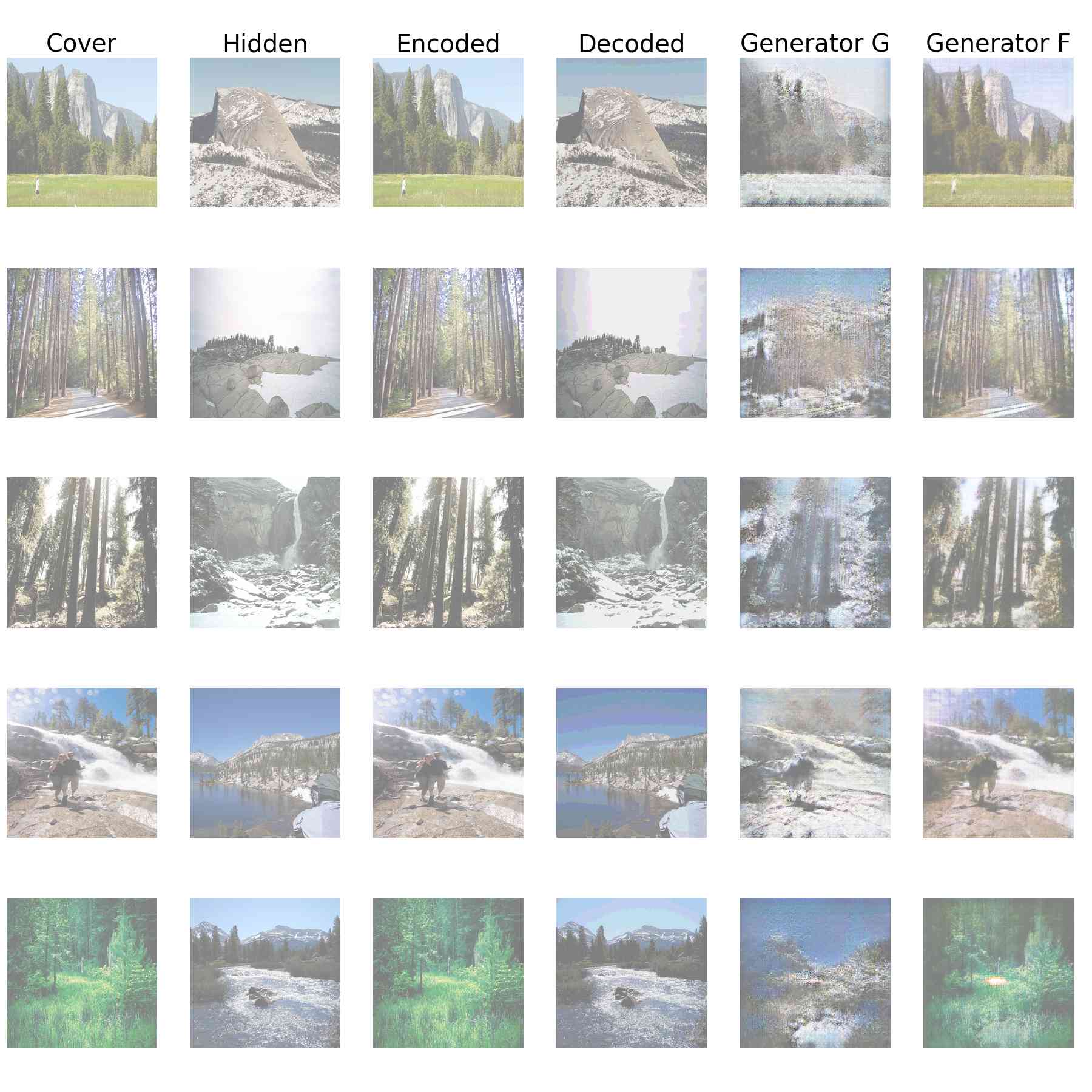}
            \caption{CycleGAN trained on 3 bits}
            \label{cycle_gan_3}
    \end{subfigure}
    \hspace{0.05\textwidth}
    \begin{subfigure}[b]{0.28\textwidth}
    \centering
            \includegraphics[scale=0.08]{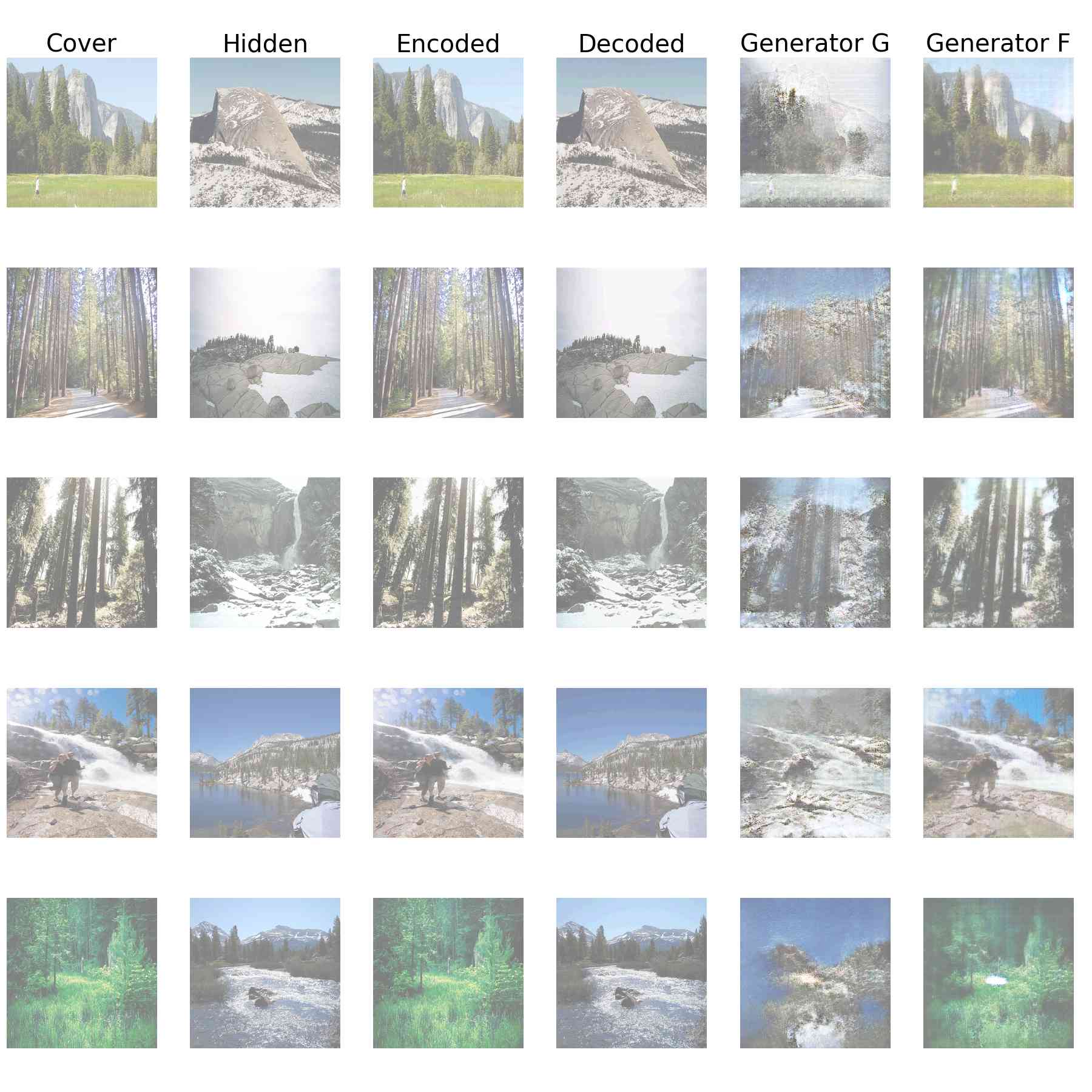}
            \caption{CycleGAN trained on 4 bits}
            \label{cycle_gan_4}
    \end{subfigure}
    \hspace{0.05\textwidth}
    \begin{subfigure}[b]{0.28\textwidth}
    \centering
            \includegraphics[scale=0.08]{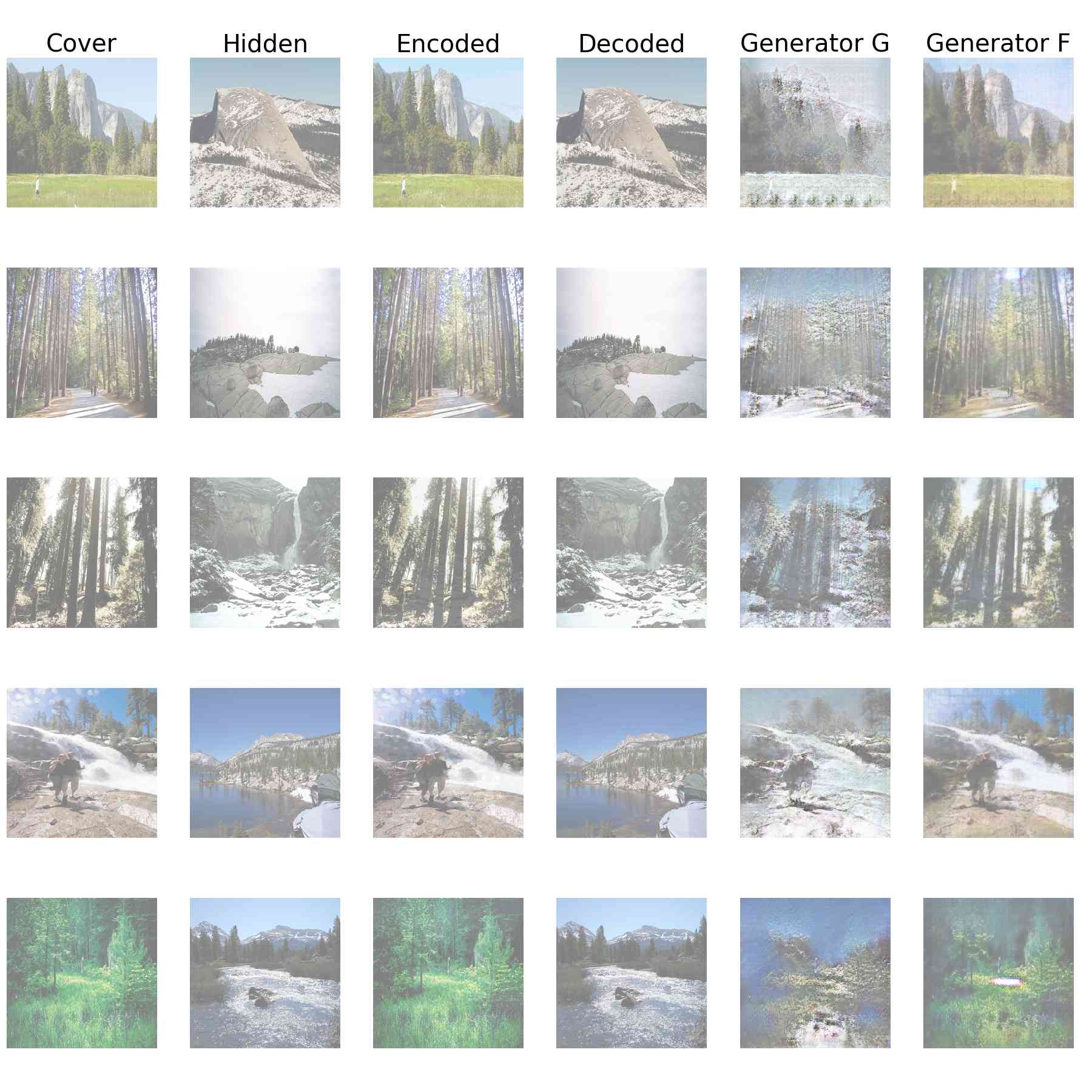}
            \caption{CycleGAN trained on 5 bits}
            \label{cycle_gan_5}
    \end{subfigure}
    \hspace{0.05\textwidth}
    \begin{subfigure}[b]{0.28\textwidth}
            \centering
                \includegraphics[ scale=0.08]{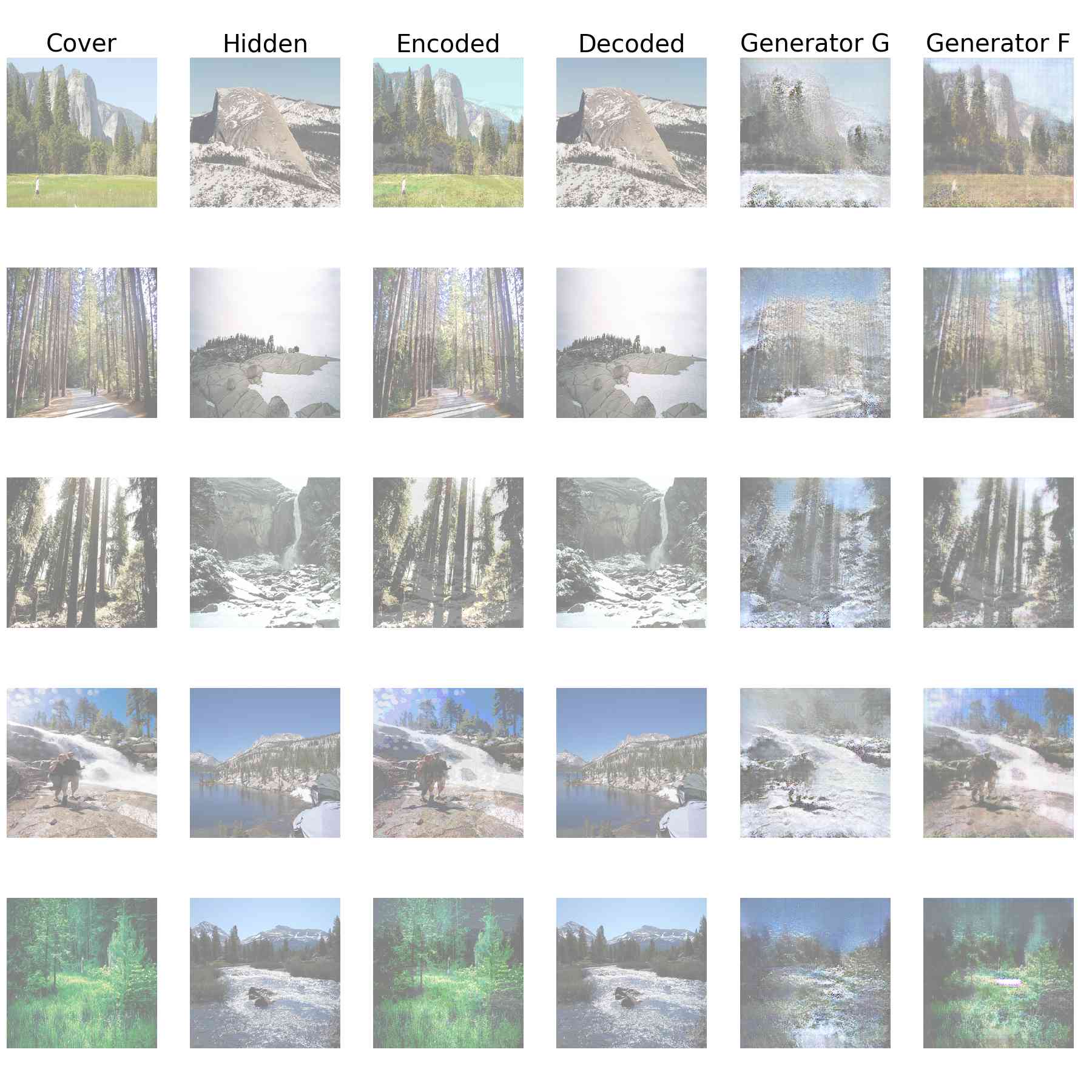}
                \caption {CycleGAN trained on 6 bits}
                \label{cycle_gan_6}
        \end{subfigure}
        \hspace{0.05\textwidth}
        \begin{subfigure}[b]{0.28\textwidth}
            \centering
                \includegraphics[scale=0.08]{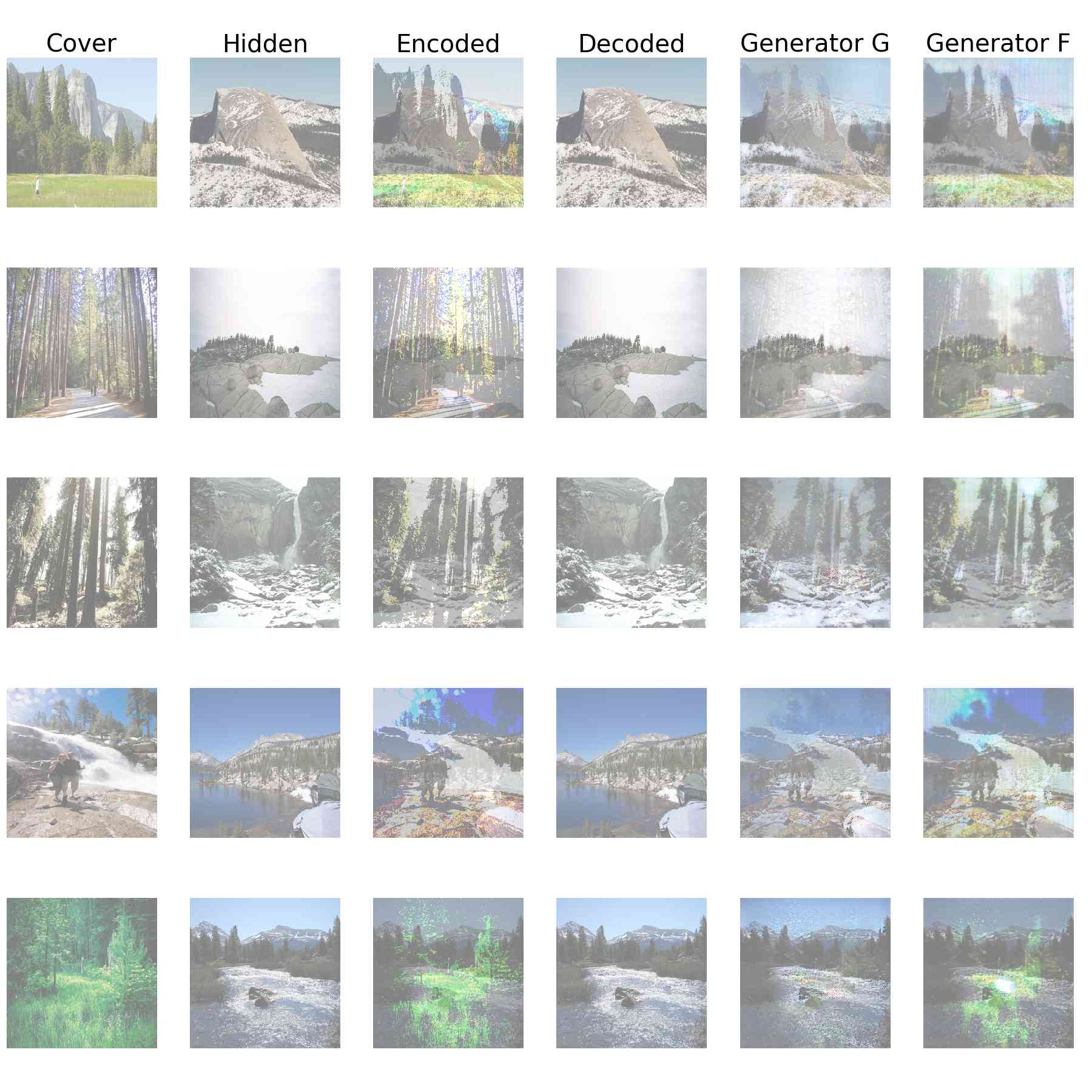}
                \caption {CycleGAN trained on 7 bits}
                \label{cycle_gan_7}
        \end{subfigure}
        \hspace{0.05\textwidth}
        \begin{subfigure}[b]{0.28\textwidth}
            \centering
                \includegraphics[ scale=0.08]{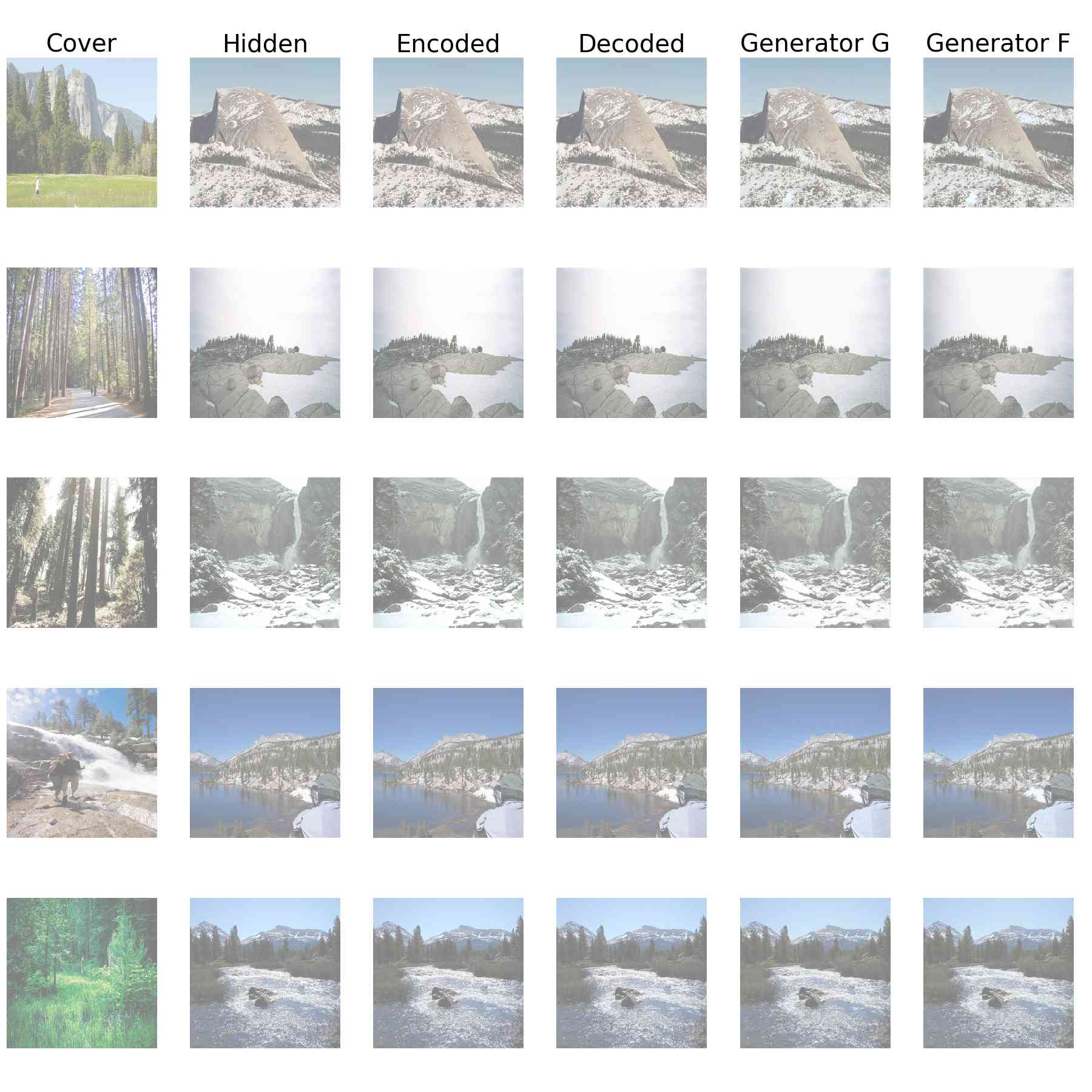}
                \caption {CycleGAN trained on 8 bits}
                \label{cycle_gan_8}
        \end{subfigure}
\caption{The images correspond to an independent CycleGAN model trained on a specific number of hidden bit sizes. Each individual column of the images corresponds to the cover images from domain one, hidden images from domain two, encoded images using that specific bit size, decoded images using that specific bit size, image generated using generator $G$ which converts from domain one to domain two, and image generated from generator $F$ which converts the previously generated image back to domain one. The sub images above are the results of predicting on five images from the training set.}

\label{fig:cycle_gan_images}
\end{figure*}

\subsubsection{Pix2pix}

Our model uses two generators and two discriminators, and these models are taken from Pix2pix as the initial starting point. The transfer of the models is possible because the Pix2pix models can be seen as pre-trained generator and discriminator models that attempt to solve a similar problem. 

Pix2pix solves the problem of paired image to image translation using Conditional GANs (cGAN). cGANs allow the generator and the discriminator to be conditioned simply by feeding the condition data, $y$, to the models \cite{mirza_conditional_2014}. This allows for the models to generate images based on the provided condition.

The objective function of the cGAN is expressed as:

\begin{align}
    \begin{split}
        \mathcal{L}_{cGAN}(G,D) &= \mathbb{E}_{x,y}[\log D(x,y)] + \\
        &= \mathbb{E}_{x,z}[\log(1-D(x,G(x,z)))]
    \end{split}
\end{align}

where $G$ tries to minimize the objective against an adversary, $D$, which tries to maximize it. Along with this objective, it has been found to be beneficial to add more traditional loss such as L2 distance \cite{pathak_context_2016}. The additional loss does not affect the discriminator, but now the generator must both fool the discriminator and be near the ground truth output in an L2 sense \cite{isola_image--image_2018}. The cGAN in Pix2pix uses L1 distance rather than L2 because it encourages less blurring:

\begin{align}
    \begin{split}
        \mathcal{L}_{L1}(G)=\mathbb{E}_{x,y,z}[{\lVert y - G(x,z) \rVert}_1]
    \end{split}
\end{align}

The final objective becomes:

\begin{align}
    \begin{split}
        G^* = {\arg}\:\underset{G}{\min}\:\underset{D}{\max}\:\mathcal{L}_{cGAN}(G,D) + \lambda\mathcal{L}_{L1}(G)
    \end{split}
\end{align}

where $\lambda$ controls the importance of the L1 distance loss. 

Due to the similarity of the two architectures, Pix2pix can be used as pre-trained models for our CycleGAN.

\begin{figure*}[!hbt]
\centering
\textbf{CycleGAN Testing}\par\medskip
    \begin{subfigure}[b]{0.28\textwidth}
    \centering
            \includegraphics[scale=0.08]{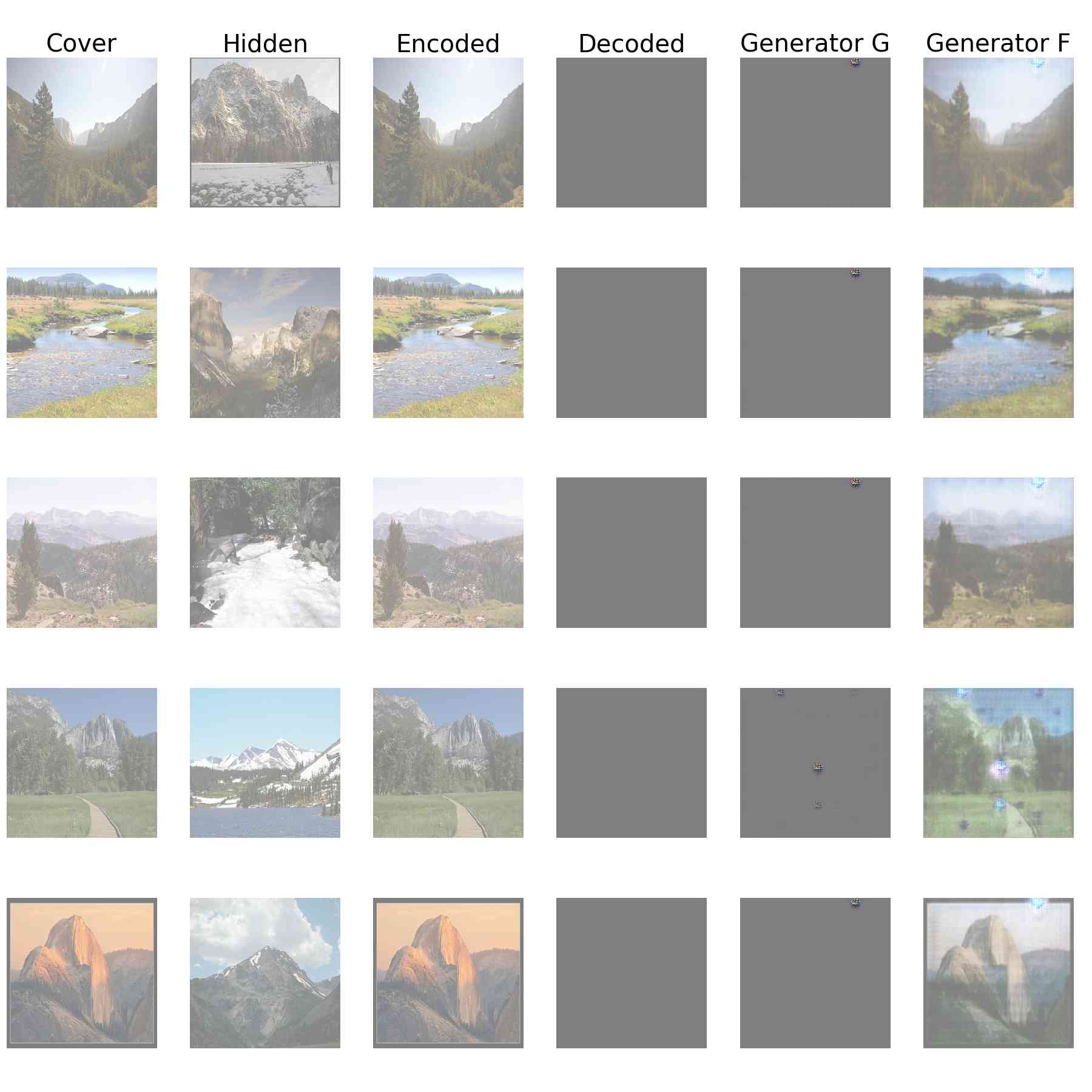}
            \caption{CycleGAN trained on 0 bits}
            \label{cycle_gan_new_0}
    \end{subfigure}
    \hspace{0.05\textwidth}
    \begin{subfigure}[b]{0.28\textwidth}
    \centering
            \includegraphics[scale=0.08]{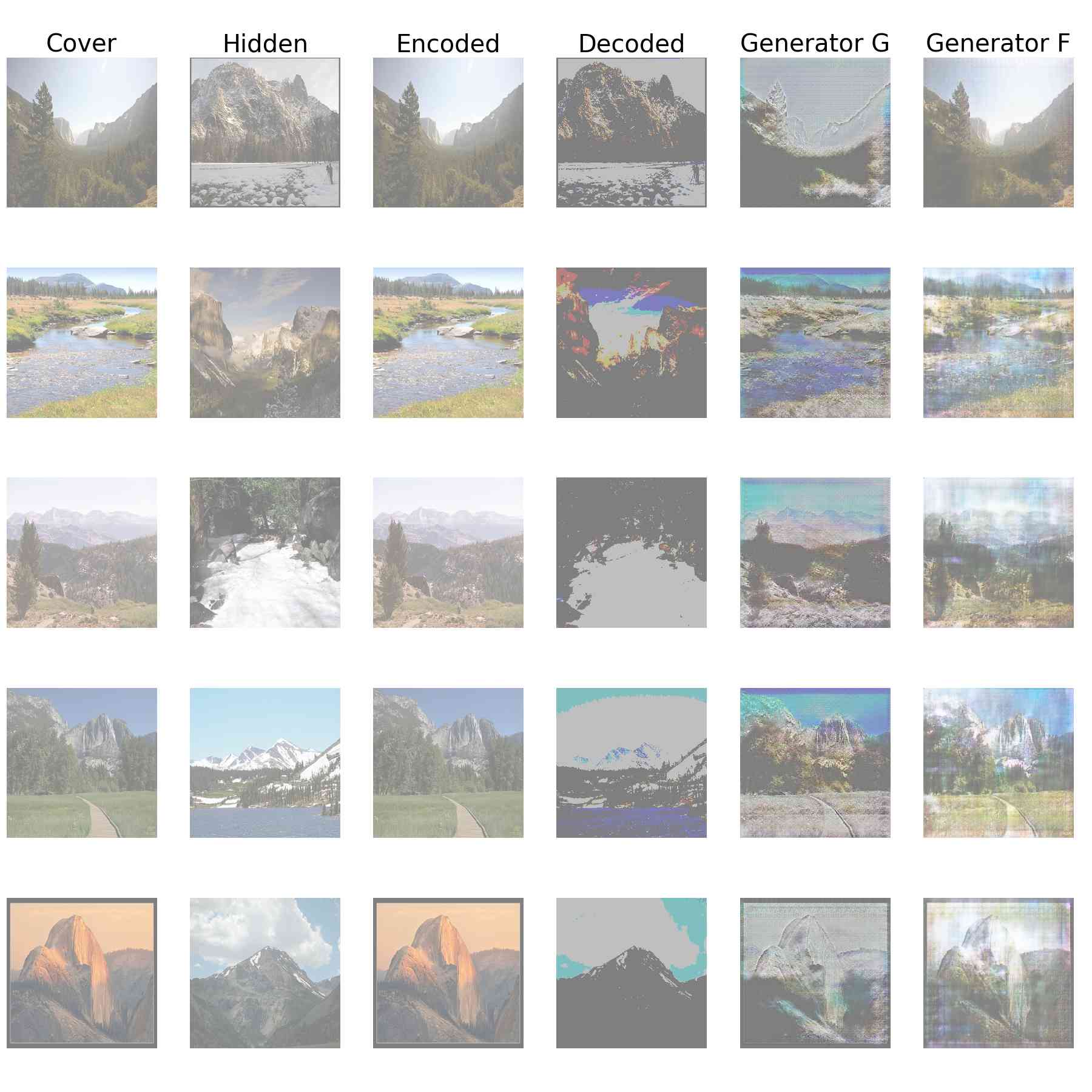}
            \caption{CycleGAN trained on 1 bit}
            \label{cycle_gan_new_1}
    \end{subfigure}
    \hspace{0.05\textwidth}
    \begin{subfigure}[b]{0.28\textwidth}
    \centering
            \includegraphics[scale=0.08]{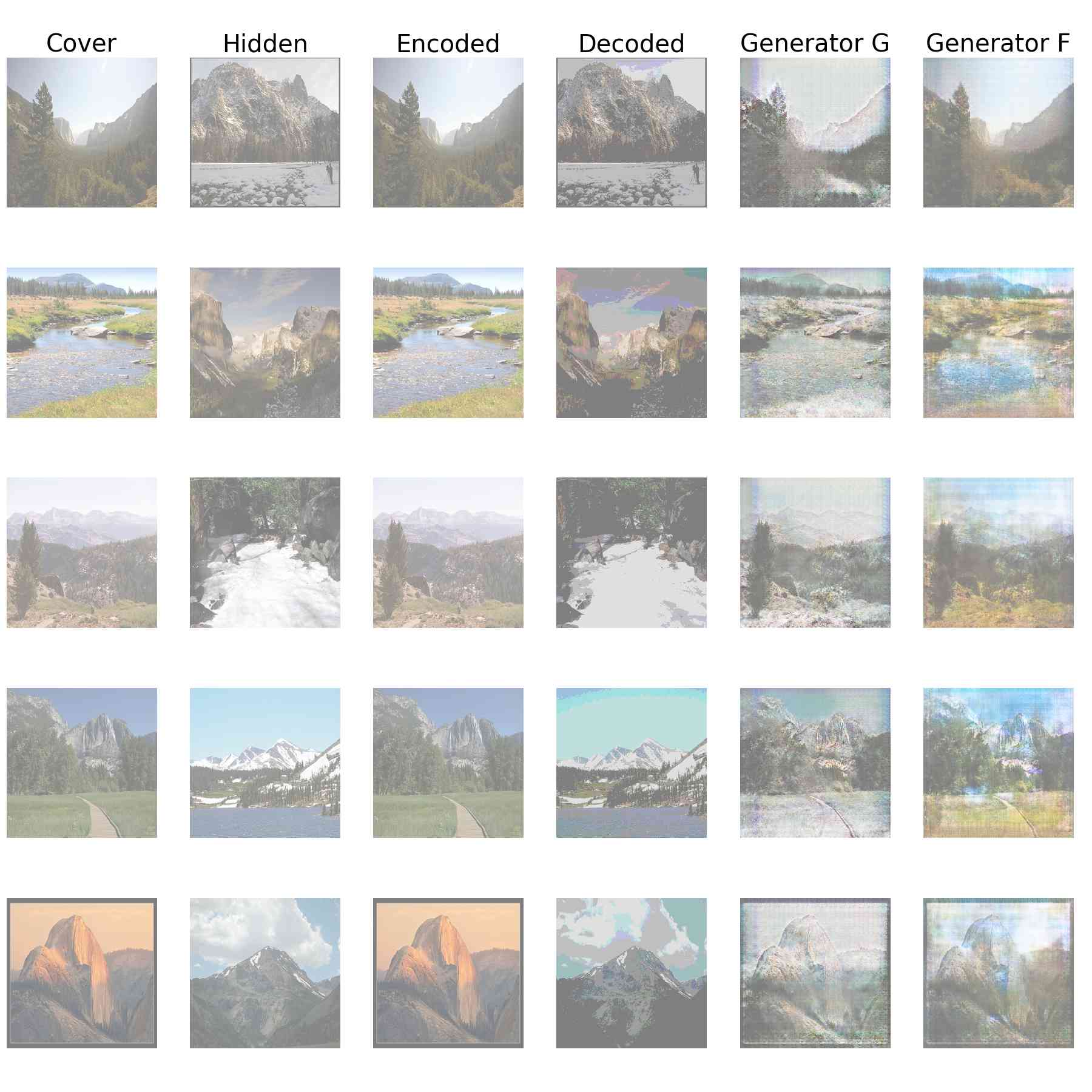}
            \caption{CycleGAN trained on 2 bits}
            \label{cycle_gan_new_2}
    \end{subfigure}
    \hspace{0.05\textwidth}
    \begin{subfigure}[b]{0.28\textwidth}
    \centering
            \includegraphics[scale=0.08]{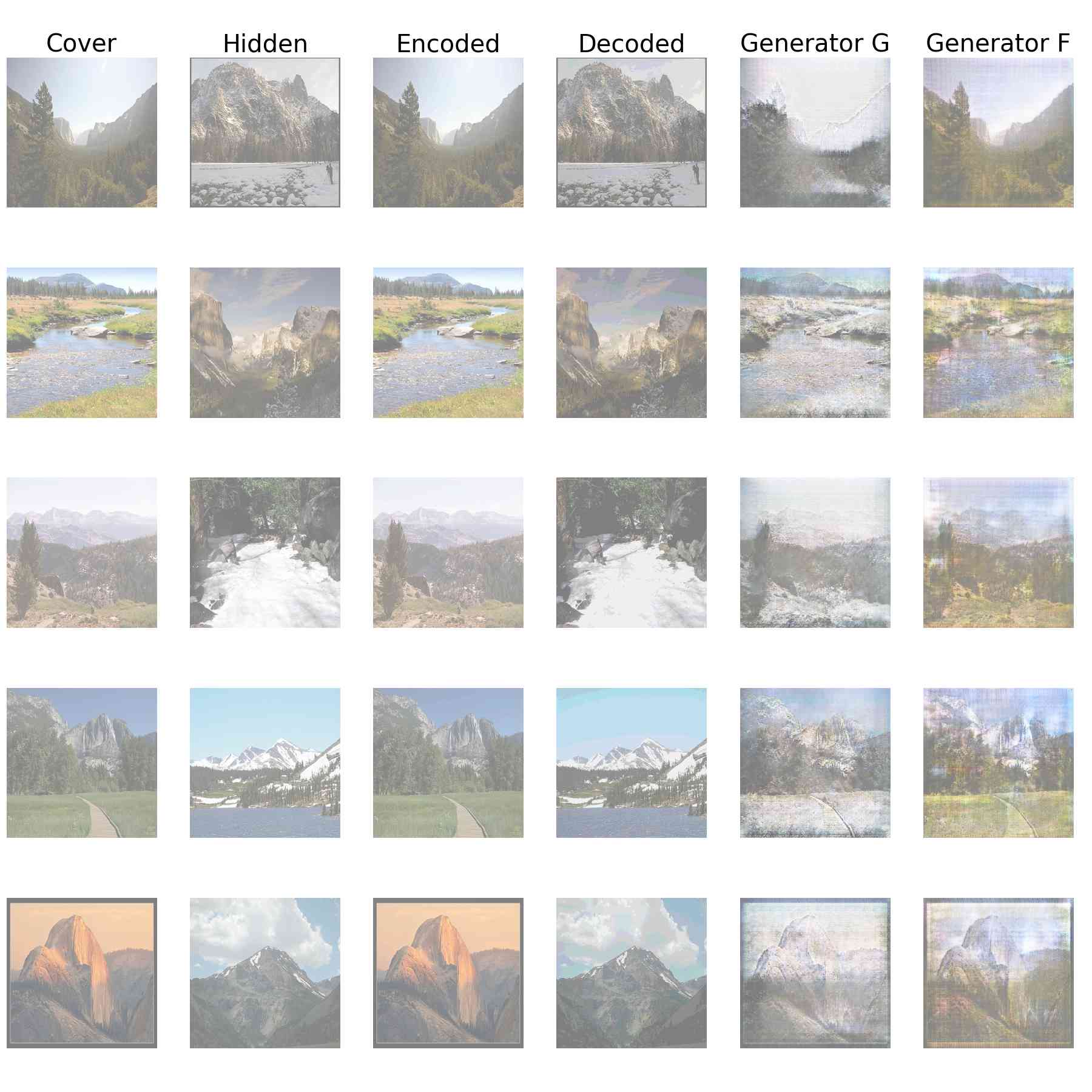}
            \caption{CycleGAN trained on 3 bits}
            \label{cycle_gan_new_3}
    \end{subfigure}
    \hspace{0.05\textwidth}
    \begin{subfigure}[b]{0.28\textwidth}
    \centering
            \includegraphics[scale=0.08]{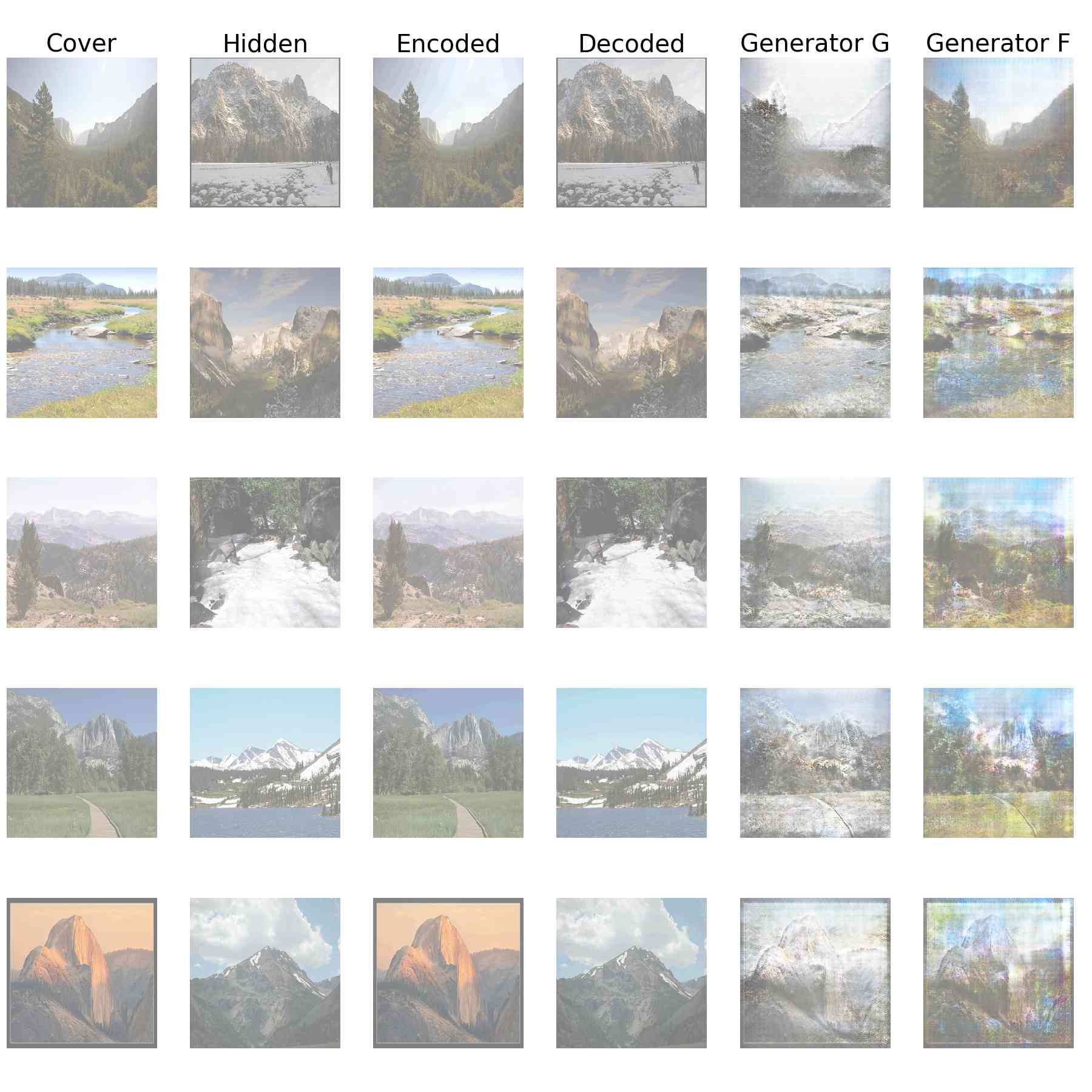}
            \caption{CycleGAN trained on 4 bits}
            \label{cycle_gan_new_4}
    \end{subfigure}
    \hspace{0.05\textwidth}
    \begin{subfigure}[b]{0.28\textwidth}
    \centering
            \includegraphics[scale=0.08]{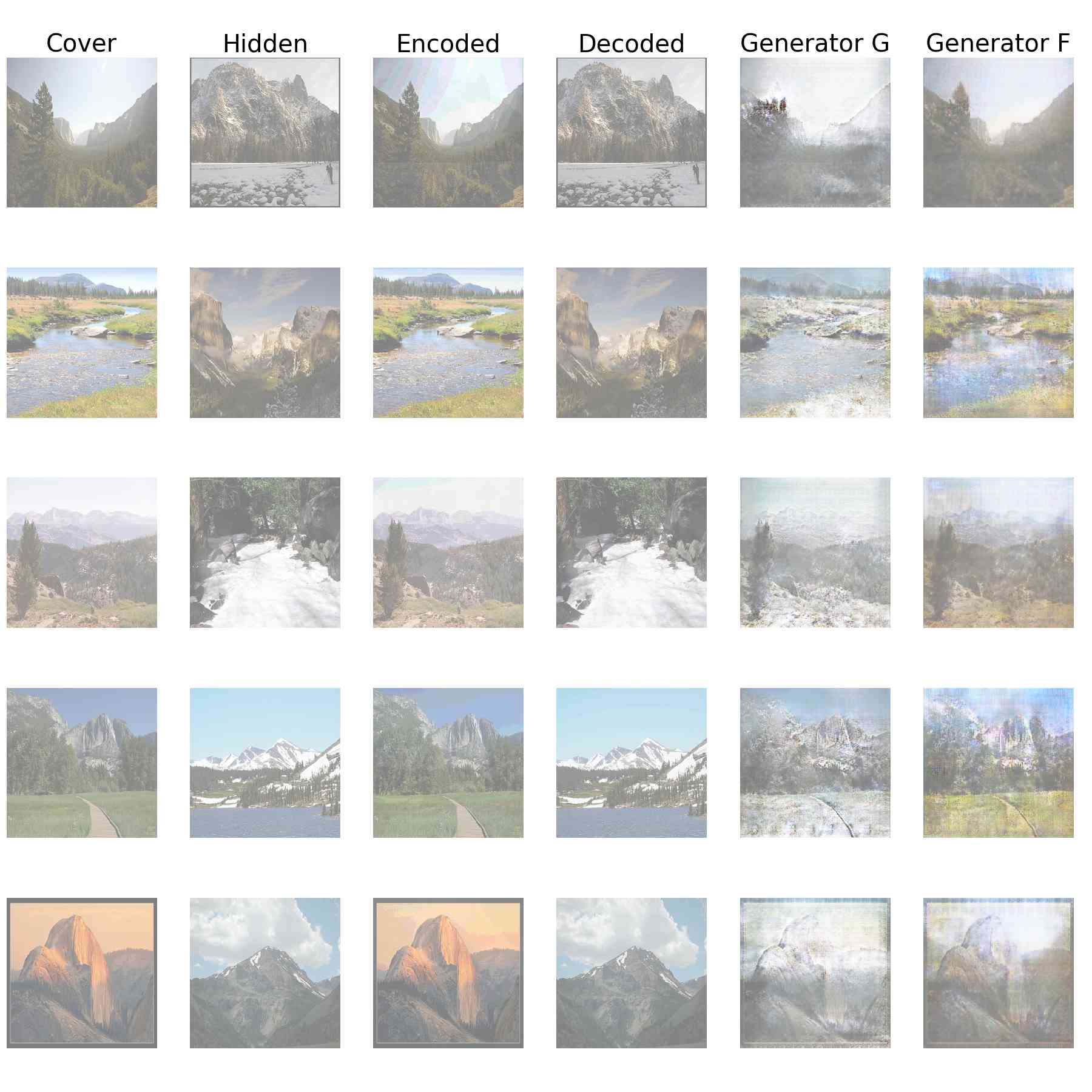}
            \caption{CycleGAN trained on 5 bits}
            \label{cycle_gan_new_5}
    \end{subfigure}
    \hspace{0.05\textwidth}
    \begin{subfigure}[b]{0.28\textwidth}
    \centering
            \includegraphics[ scale=0.08]{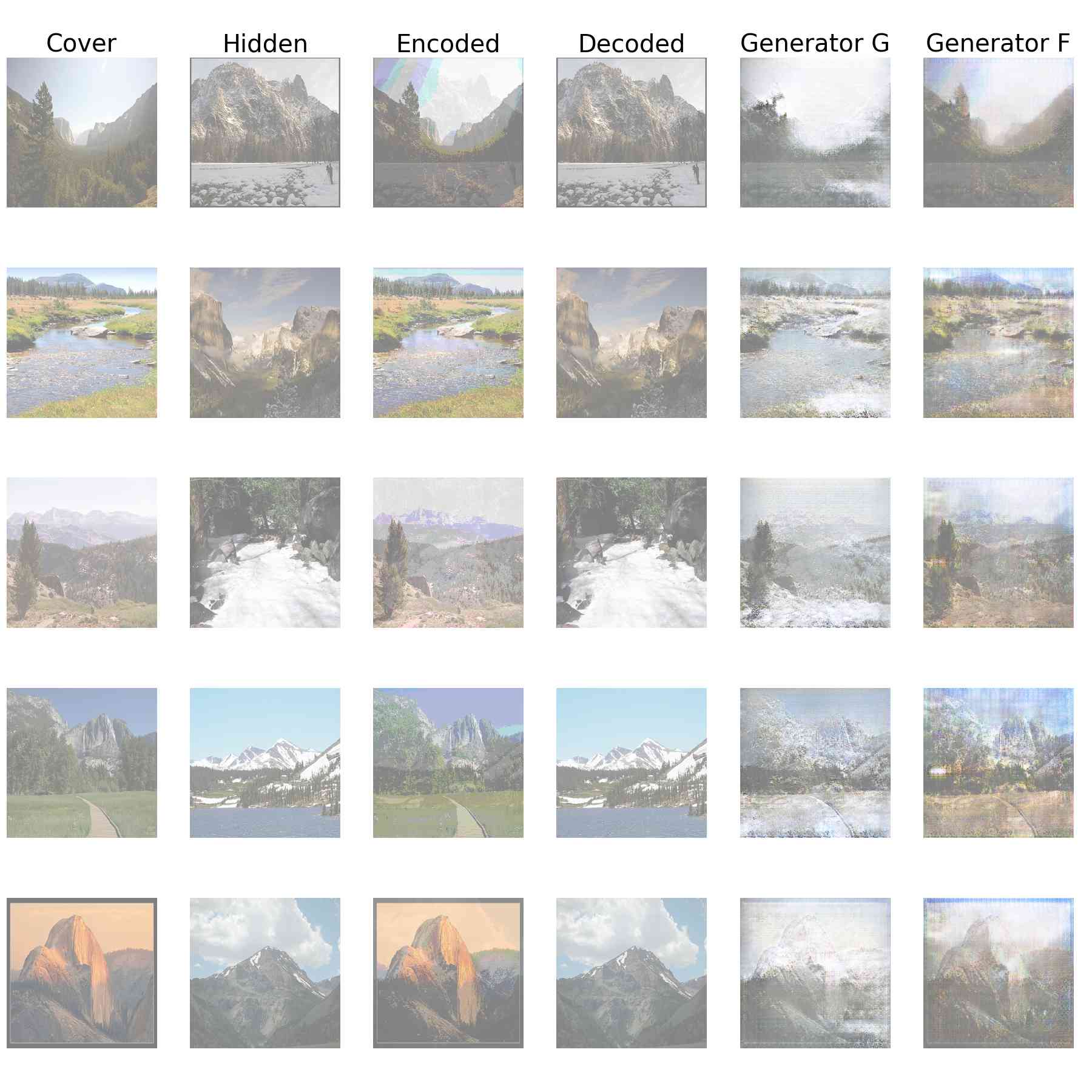}
            \caption {CycleGAN trained on 6 bits}
            \label{cycle_gan_new_6}
    \end{subfigure}
    \hspace{0.05\textwidth}
    \begin{subfigure}[b]{0.28\textwidth}
        \centering
            \includegraphics[scale=0.08]{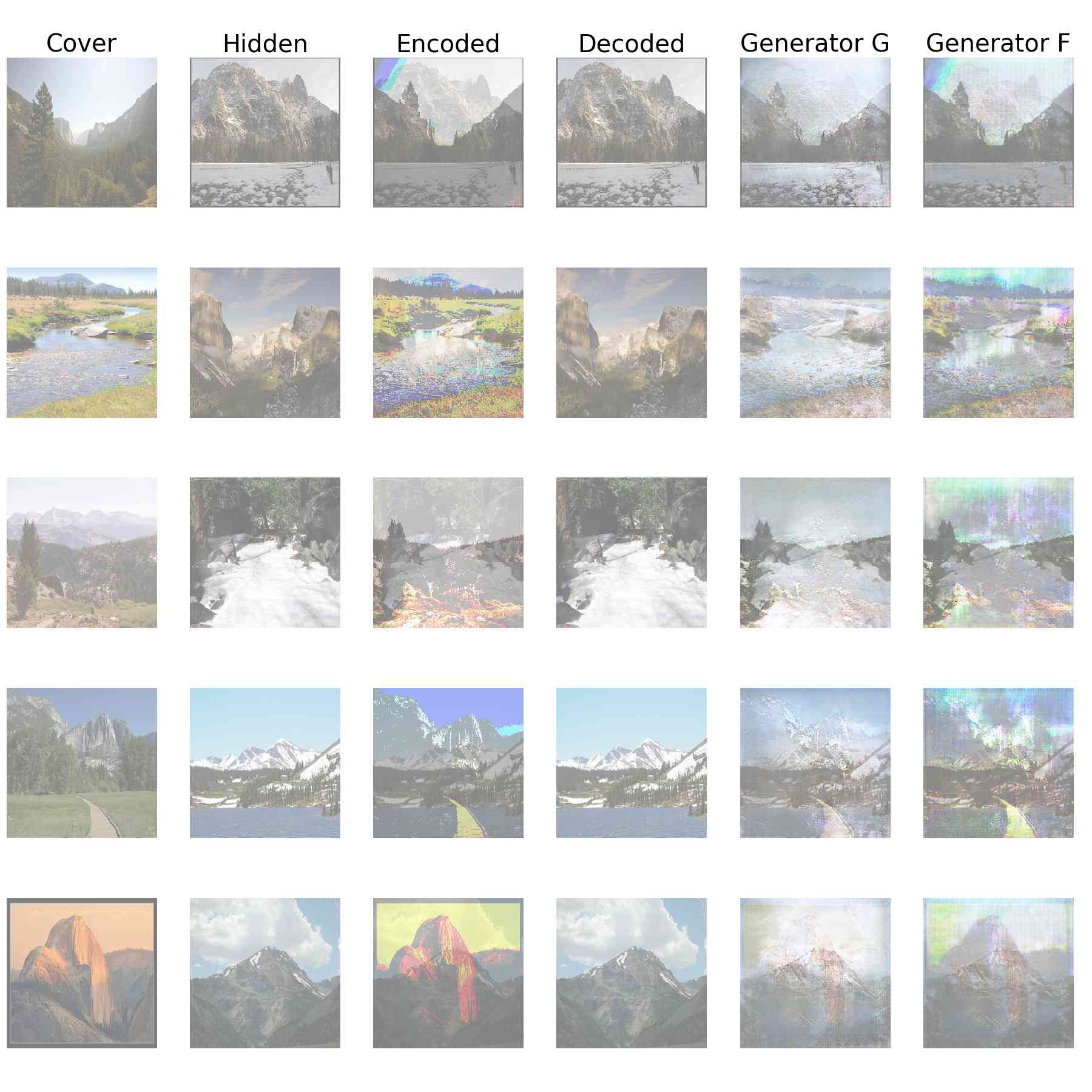}
            \caption {CycleGAN trained on 7 bits}
            \label{cycle_gan_new_7}
    \end{subfigure}
    \hspace{0.05\textwidth}
    \begin{subfigure}[b]{0.28\textwidth}
        \centering
            \includegraphics[ scale=0.08]{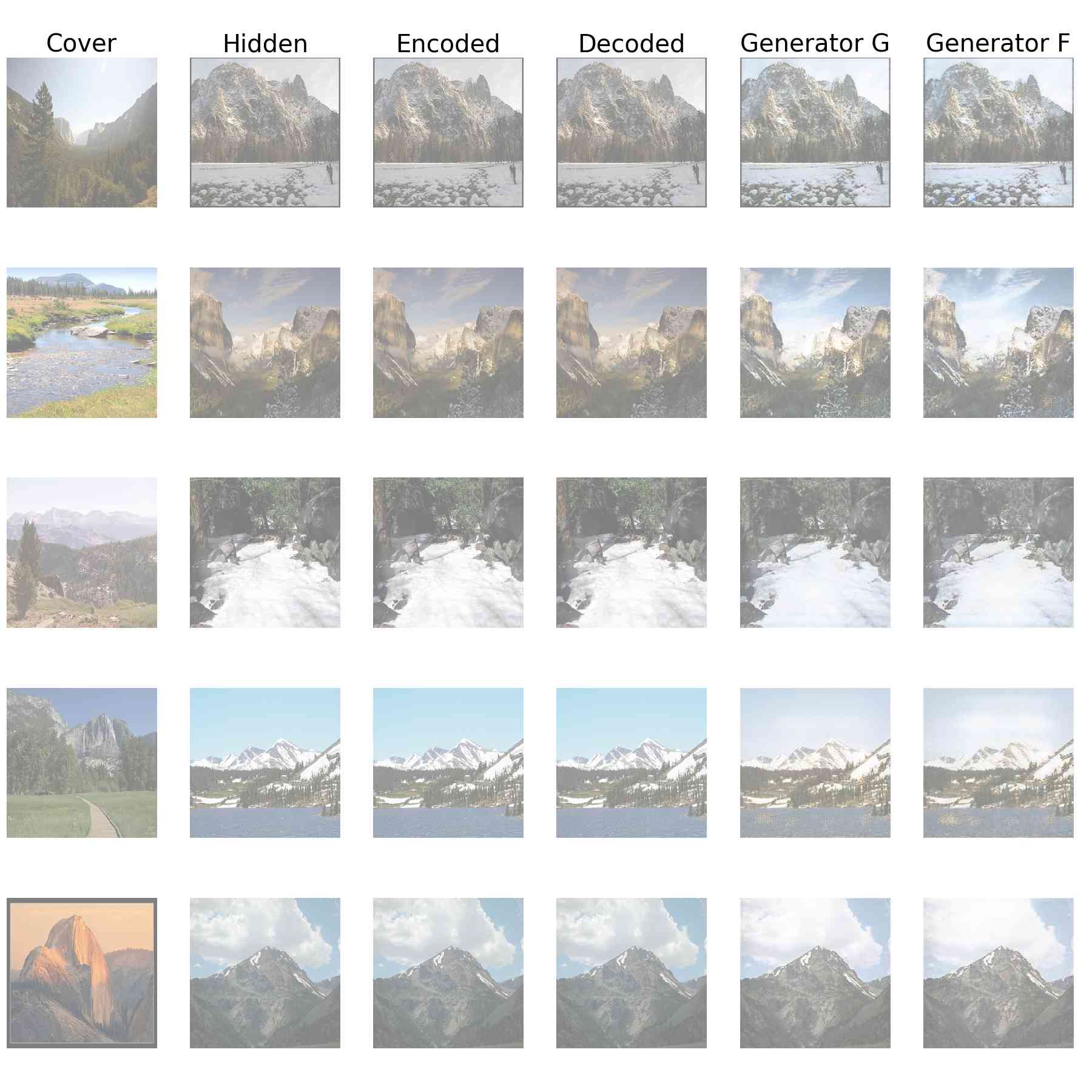}
            \caption {CycleGAN trained on 8 bits}
            \label{cycle_gan_new_8}
    \end{subfigure}
\caption{Each of the images corresponds to an independent CycleGAN model trained on a specific number of hidden bit sizes. Each individual column of the images corresponds to the cover images from domain one, hidden images from domain two, encoded images using that specific bit size, decoded images using that specific bit size, and the predicted decoded image by the CycleGAN, respectively. The CycleGAN trys to predict five images from its testing set which it has never seen before.}
\label{fig:cycle_gan_new_images}
\end{figure*}

\begin{figure*}[!hbt]
\centering
\textbf{Bayesian Optimization Results}\par\medskip
    \begin{subfigure}[b]{0.28\textwidth}
    \centering
            \includegraphics[scale=0.08,clip=false]{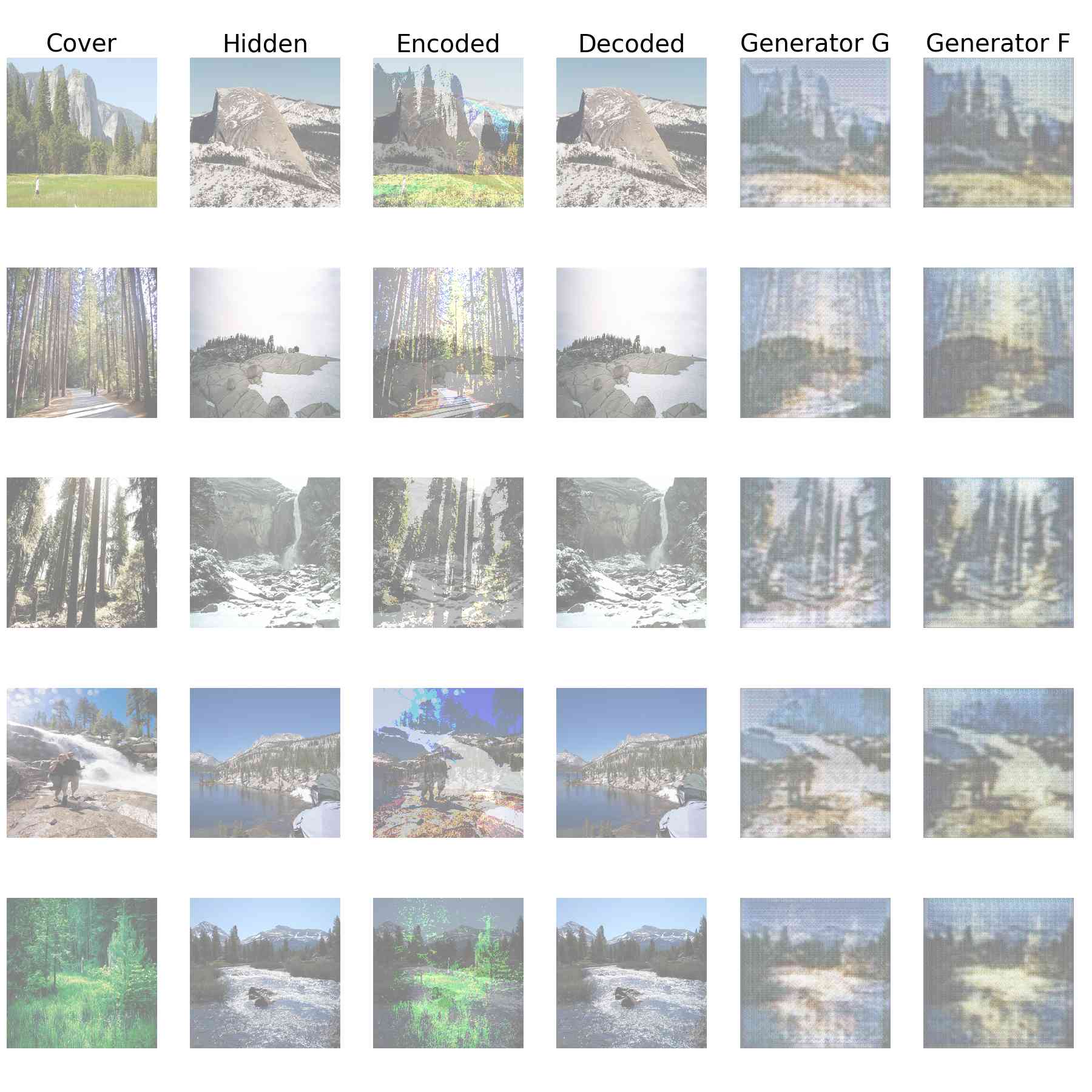}
            \caption{Before Bayesian Optimization}
            \label{Bayes_Bad}
    \end{subfigure}
    \hspace{0.05\textwidth}
    \begin{subfigure}[b]{0.28\textwidth}
    \centering
            \includegraphics[scale=0.08,clip=false]{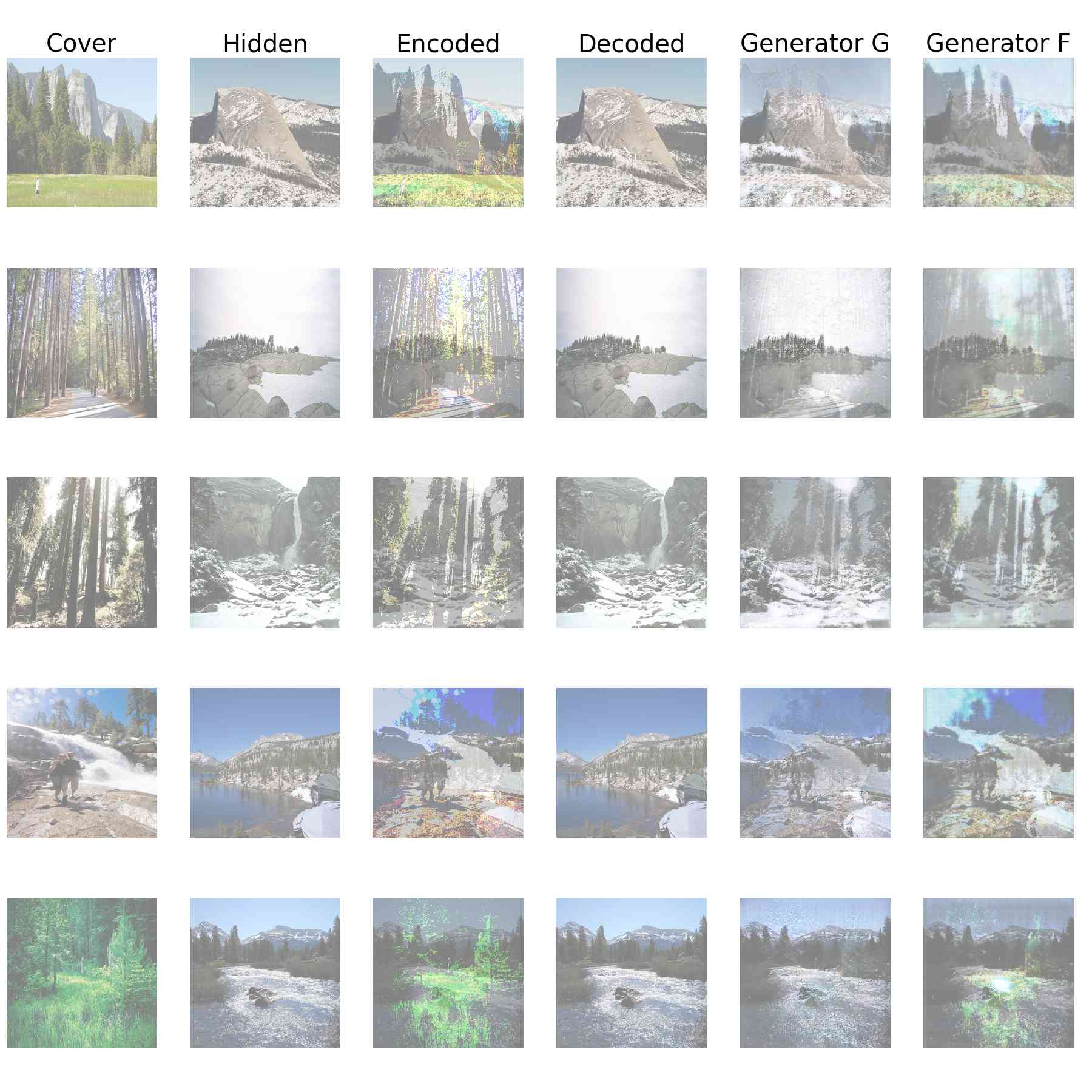}
            \caption{After Bayesian Optimization}
            \label{Bayes_Good}
    \end{subfigure}
    \hspace{0.05\textwidth}
\caption{The sub-images above show the results of Bayesian Optimization. The first sub-image shows the accuracy of the model before running Bayesian Optimization, and the second image shows the accuracy of the model after running Bayesian Optimization.}
\label{fig:bayes_opt_images}
\end{figure*}

\subsection{Autoencoders Model Description}
We implemented a Convolutional Autoencoder to test the accuracy of our CycleGAN against. The Convolutional Autoencoder follows the same encoder and decoder method of the original Autoencoder \cite{ng2011sparse}, but it implements Convolutional Neural Networks (CNN) for the encoder and decoder \cite{chen_deep_2017}. Our implementation modifies the traditional input and output of the Autoencoder to work with the encoded and decoded images.

The encoder takes the encoded image as the input and uses a deep CNN to reduce its dimensionality, then the decoder converts the latent representation of the encoded image to generate the decoded image. 

With this implementation, we have a model to compare the success of our CycleGAN against. 

\begin{figure*}[!hbt]
\centering
\textbf{Training CycleGAN with Varying Bit Sizes}\par\medskip
    \begin{subfigure}[b]{0.27\textwidth}
    \centering
            \includegraphics[scale=0.07]{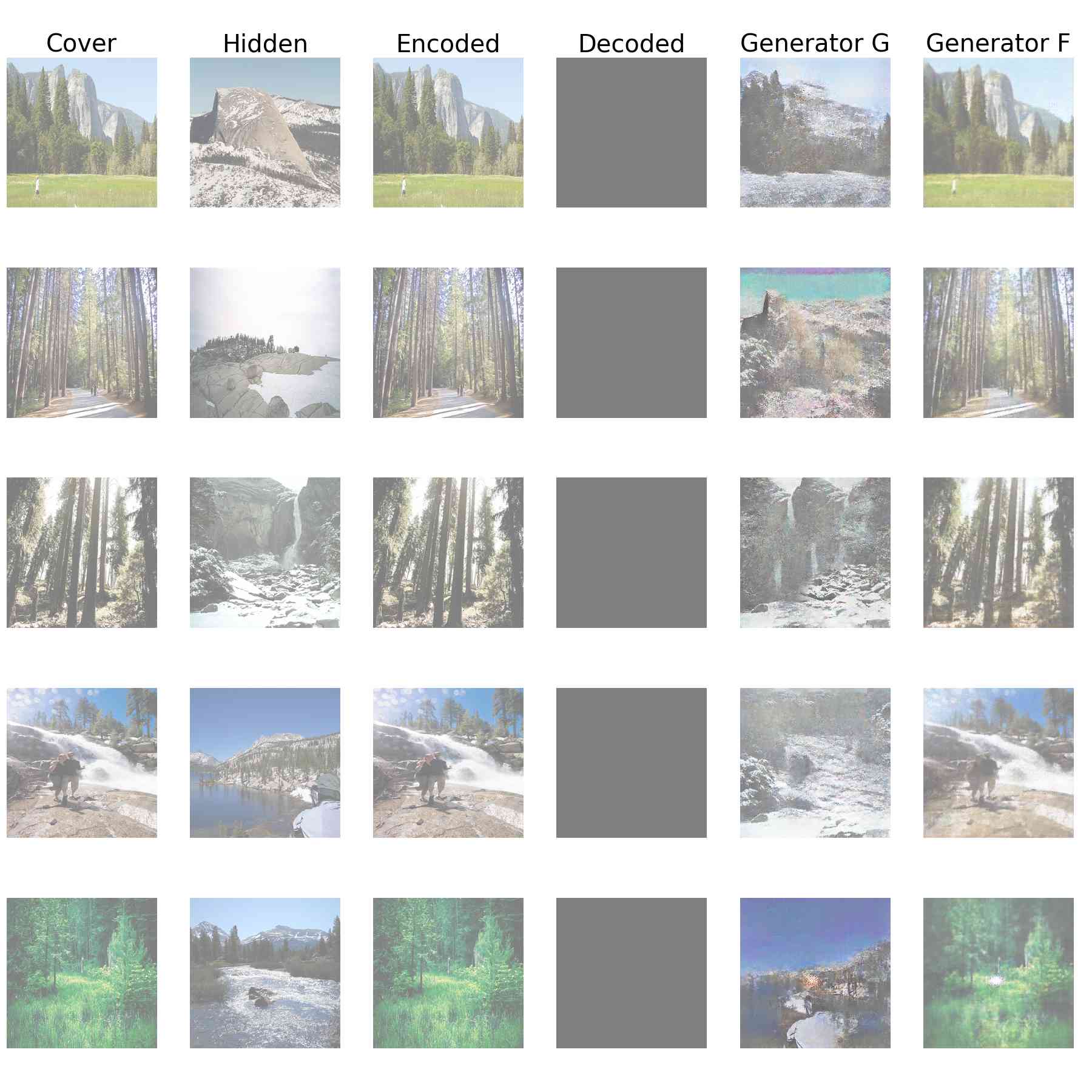}
            \caption{Generated images with bit size 0}
            \label{bit_size_0}
    \end{subfigure}
    \hspace{0.05\textwidth}
    \begin{subfigure}[b]{0.27\textwidth}
    \centering
            \includegraphics[scale=0.07]{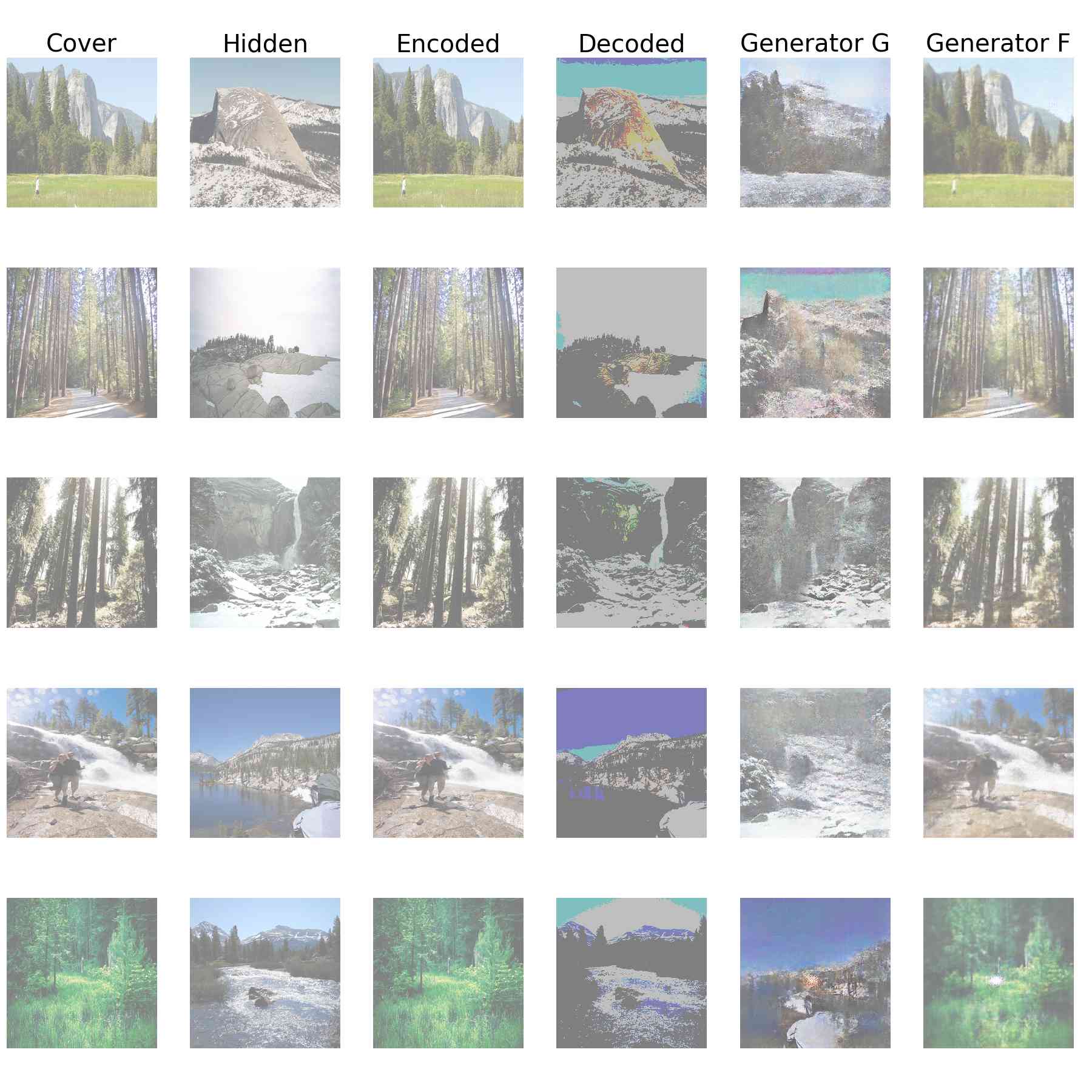}
            \caption{Generated images with bit size 1}
            \label{bit_size_1}
    \end{subfigure}
    \hspace{0.05\textwidth}
    \begin{subfigure}[b]{0.27\textwidth}
    \centering
            \includegraphics[scale=0.07]{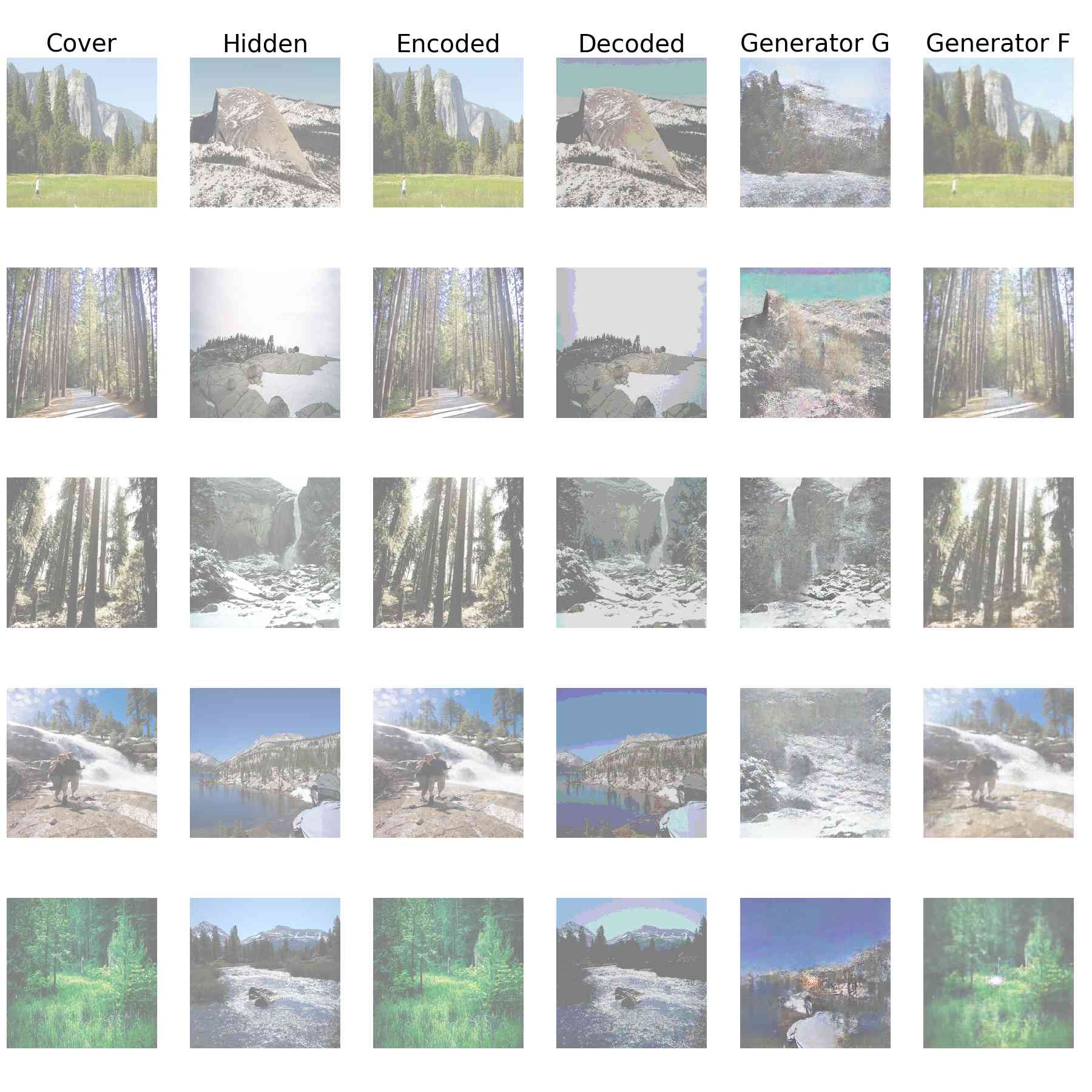}
            \caption{Generated images with bit size 2}
            \label{bit_size_2}
    \end{subfigure}
    \hspace{0.05\textwidth}
    \begin{subfigure}[b]{0.27\textwidth}
    \centering
            \includegraphics[scale=0.07]{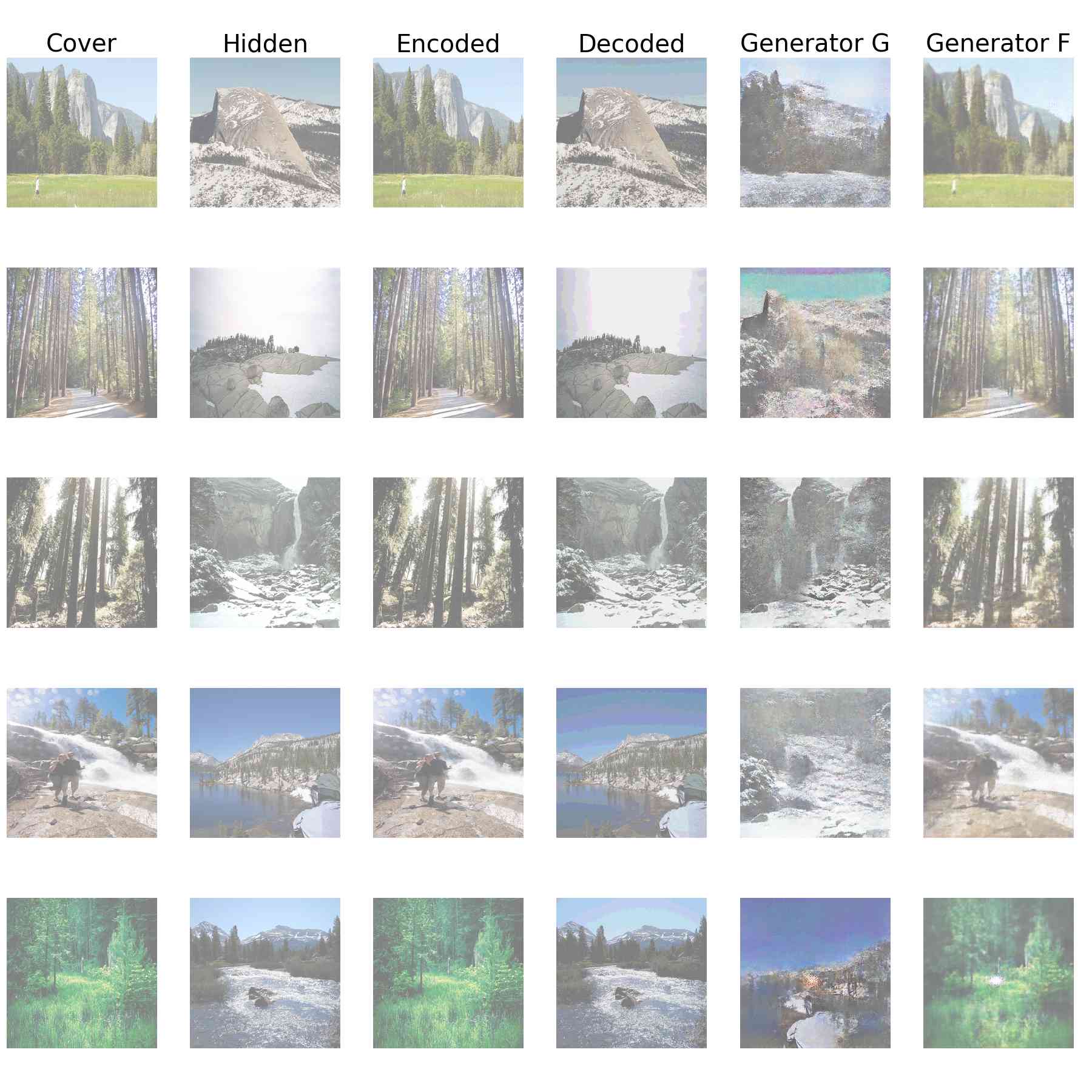}
            \caption{Generated images with bit size 3}
            \label{bit_size_3}
    \end{subfigure}
    \hspace{0.05\textwidth}
    \begin{subfigure}[b]{0.27\textwidth}
    \centering
            \includegraphics[scale=0.07]{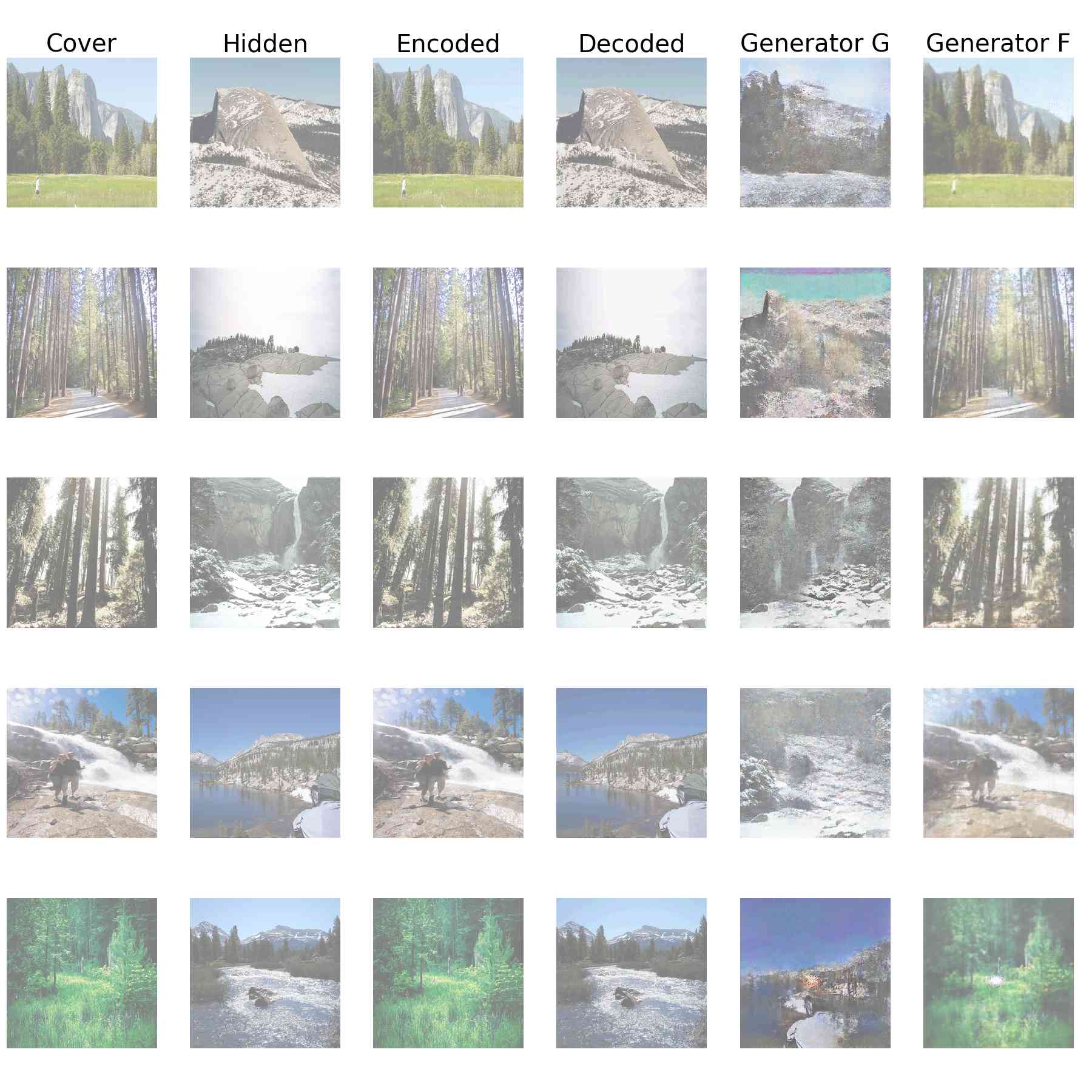}
            \caption{Generated images with bit size 4}
            \label{bit_size_4}
    \end{subfigure}
    \hspace{0.05\textwidth}
    \begin{subfigure}[b]{0.27\textwidth}
    \centering
            \includegraphics[scale=0.07]{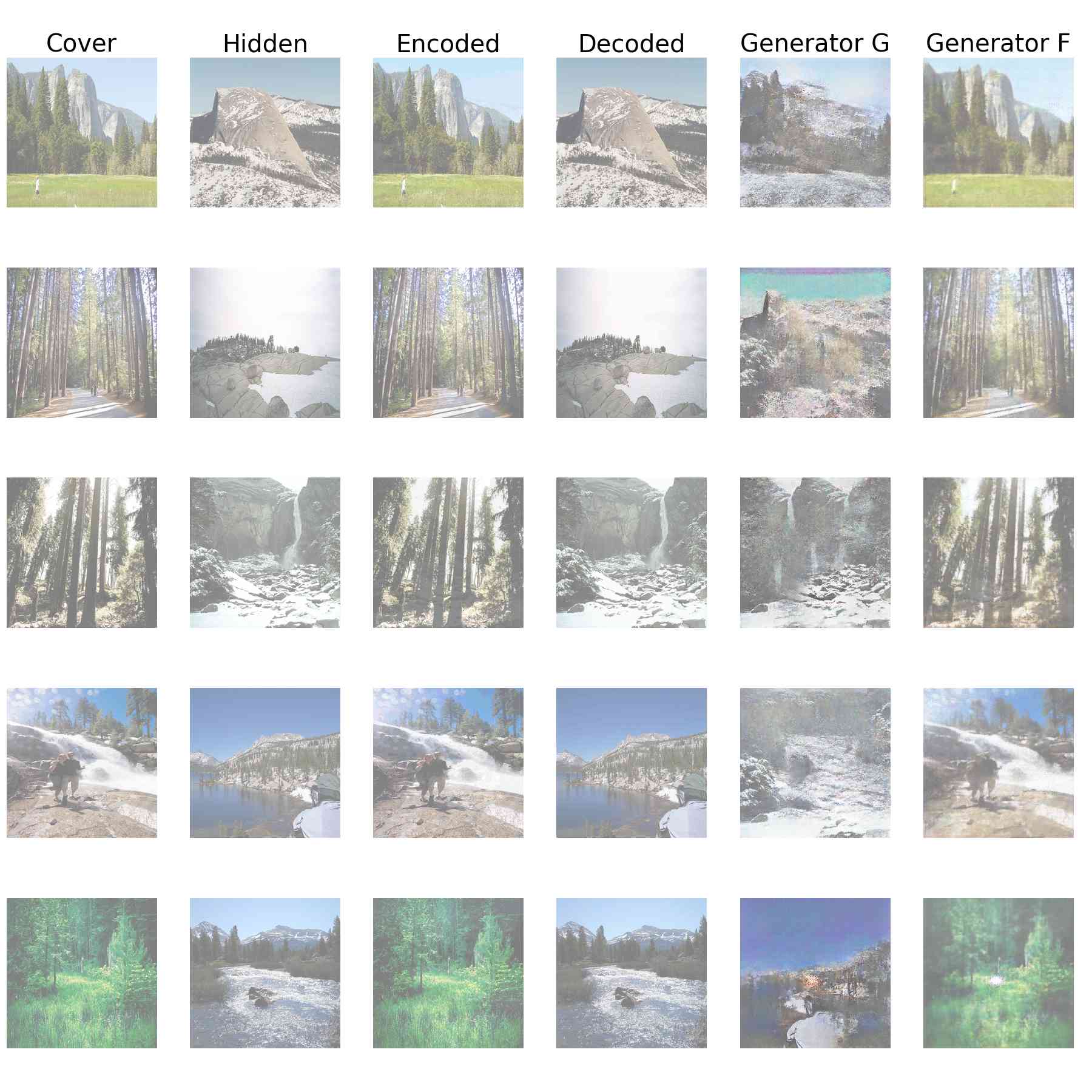}
            \caption{Generated images with bit size 5}
            \label{bit_size_5}
    \end{subfigure}
    \hspace{0.05\textwidth}
    \begin{subfigure}[b]{0.27\textwidth}
            \centering
                \includegraphics[ scale=0.07]{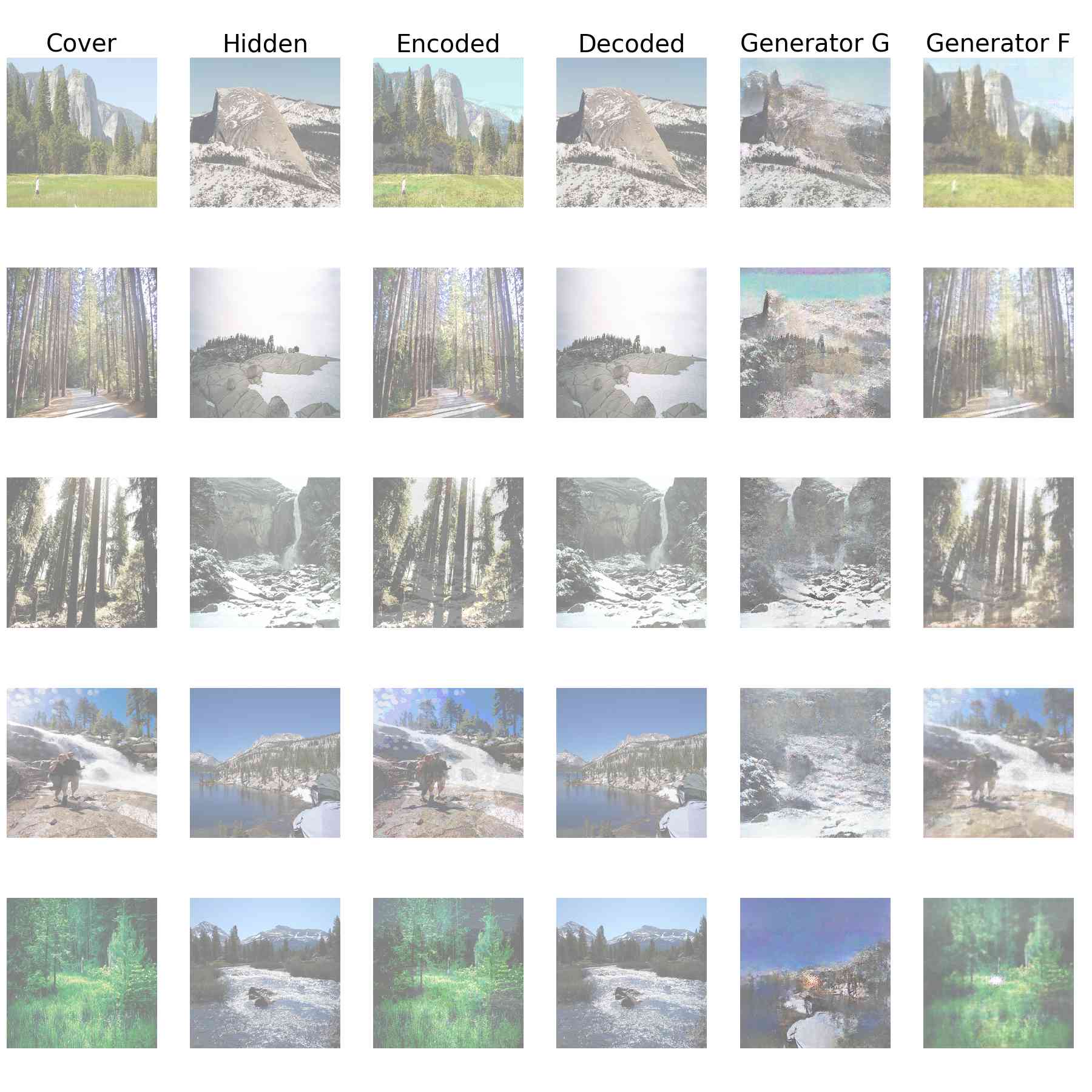}
                \caption {Generated images with bit size 6}
                \label{bit_size_6}
        \end{subfigure}
        \hspace{0.05\textwidth}
        \begin{subfigure}[b]{0.27\textwidth}
            \centering
                \includegraphics[scale=0.07]{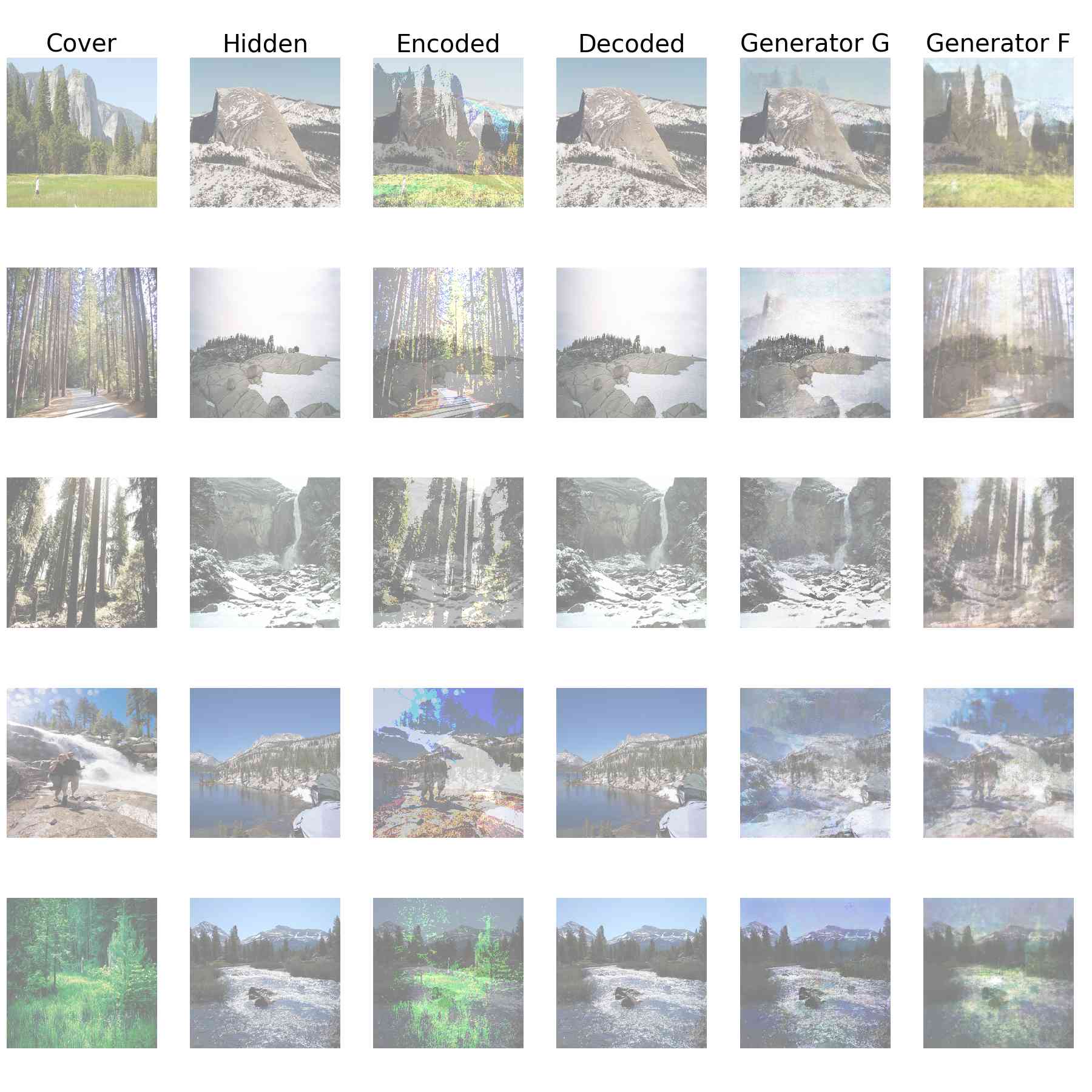}
                \caption {Generated images with bit size 7}
                \label{bit_size_7}
        \end{subfigure}
        \hspace{0.05\textwidth}
        \begin{subfigure}[b]{0.27\textwidth}
            \centering
                \includegraphics[ scale=0.07]{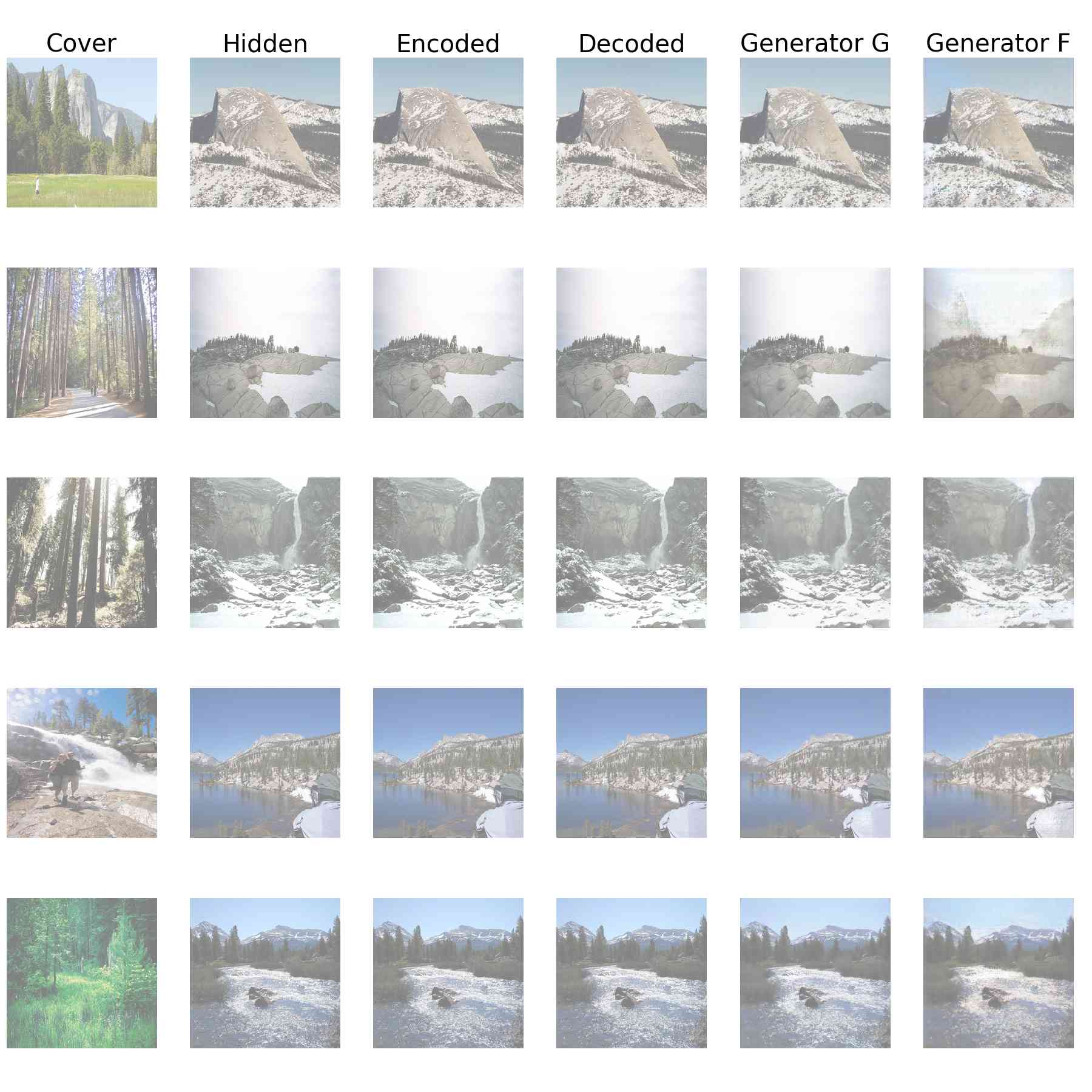}
                \caption {Generated images with bit size 8}
                \label{bit_size_8}
        \end{subfigure}
\caption{A single CycleGAN model trained with varying bit sizes and sub images above show the results of five images of its training domain. Each individual column of the images corresponds to the cover images from domain one, hidden images from domain two, encoded images using that specific bit size, decoded images using that specific bit size, and the predicted decoded image by the CycleGAN, respectively.}
\label{fig:bit_size_images}
\end{figure*}

\subsection{Testing Protocols}
Different training and testing protocols were introduced for the CycleGAN and Autoencoder models. On top of the individual testing protocols, we implemented Bayesian Optimization to improve the accuracy of our model by introducing hyper-parameter tuning.  

\subsubsection{Bayesian Optimization}
Bayesian Optimization assumes that the noise in a task lies in the hyper-parameters rather than the formulation of the problem. Therefore, it tries to improve the model not by changing the formulation of the problem but rather by tuning the hyper-parameters using Bayesian Inference.

Bayesian optimization forms a posterior distribution of functions that best describes the black-box function that needs to be optimized (in this case, our CycleGAN model). As Bayesian optimization continues to observe more and more, the posterior distribution improves, and the algorithm learns which regions of the hyper-parameter search space are worth exploring and which are not.

While the algorithm iterates over its search space, it balances exploration and exploitation by taking into account what it knows about the black-box function. At each iteration, a Gaussian Process \cite{rasmussen_gaussian_2006} is fitted to the points it has already explored, and the posterior distribution, along with an exploration strategy, is used to determine the optimal next step \cite{bayes_opt, brochu_tutorial_2010, snoek_practical_2012}. 

This algorithm aims to minimize the number of steps needed to find the optimal combination of hyper-paramters. To accomplish this, the algorithm uses a proxy optimization problem that is cheaper and solvable with common tools. We implemented Bayesian optimization as it is most adequate in situations where sampling the black-box function is expensive. 

\subsubsection{Cycle Generative Adversarial Network Training and Testing}
The RGB values for images are set using 8 bits of information. This number means that our LSB algorithm can sample anywhere from 0-8 bits of information to be encoded in the cover file. Therefore, all 9 bit sizes were used to train and test our CycleGAN algorithm.

For each possible bit size, we implemented an individual CycleGAN model that is trained only on that particular bit size and tested on the same size. We implemented a total of 9 CycleGAN models, each on a specific bit size. To test the models on each bit size, we use the 9 CycleGAN models and give them previously unseen encoded images created with their respective bit size.

Because tuning hyper-parameters is time expensive, we were only able to use Bayesian optimization on a single CycleGAN model.
    
\subsubsection{Autoencoder Training and Testing}
Just as with the CycleGAN models, we implemented 9 Autoencoder models to test their capabilities at each number of bit sizes. The training and testing methods for the Autoencoders are the same as the CycleGAN models, except for the respective models used.

\subsection{Bit Size Variation}
We implemented a technique to use one CycleGAN model to account for all possible bit sizes, instead of having to create a unique model for each size. This was done by training a single CycleGAN model, but at each training epoch, the bit size used by the LSB algorithm was randomly chosen between 0-9 bits. The model is then trained for a certain number of steps on that particular bit size.
    
\section{Results}

By analyzing Figure \ref{fig:autoencoder_images}, we can see the results of Autoencoders attempting to crack the Steganography algorithm. Each sub image in Figure \ref{fig:autoencoder_images} corresponds to an Autoencoder trained on a specific number of hidden bit sizes in the Steganography algorithm. Each Autoencoder is tested by trying to predict the decoded image by using the encoded image as the input. The success of the model can be analyzed by looking at the predicted column of each sub image. 

The predicted images by an Autoencoder trained on zero to four bits have little to no validity and what is being produced is mainly noise. For the rest of the bit sizes, the predicted images seem more clear. However, each predicted image is the same, which is a clear indication of failure. Furthermore, the repetitively predicated images hint to the Autoencoders memorizing rather than generalizing. Even when the encoded image contains all the information, such as with 8 hidden bits, the Autoencoder is still unable to generate the appropriate image.

Figure \ref{fig:cycle_gan_images} shows the results of our CycleGAN algorithm on the training data. The success of our model can be analyzed by referring to generator $G$ column in each of the sub images. This column contains the results of converting from domain one, encoded images, to domain two, decoded images. The cycle consistency of our algorithm can be analyzed by referring to the generator $F$ column which takes the generated image from generator $G$ and converts it back to domain one.

From the Figure \ref{fig:cycle_gan_images}, it is clear that the model is able to start predicting images with just one hidden bit size available to it. As the number of hidden bit sizes increase, the model is able to predict with more accuracy. When the model has six to eight hidden bits available to it, it can predict with significant accuracy. Furthermore, each predicted image is unique, which hints to the model generalizing rather than memorizing. 

To test the accuracy of CycleGAN on data it has never seen before we created a set of images using cover and hidden images that the CycleGAN was not trained on. The results are seen in Figure \ref{fig:cycle_gan_new_images}. The Figure clearly shows the ability of the CycleGAN to decrypt images it has never seen before. While it is clear that the algorithm is not perfect, it does show clear signs of success.

To improve the accuracy of the model, we implemented Bayesian optimization. The results of running Bayesian optimization is shown in Figure \ref{fig:bayes_opt_images}, and its success is clear. 

An additional training technique, varying the bit size, was implemented, and the results of this technique is shown in Figure\ref{fig:bit_size_images}. The effectiveness of varying bit size is minimal, but it allows for a single model to be trained in all bit sizes.

\section{Discussion}
Steganography is the art of concealing a message within another message. In this case, images are being used as our messages. There is a high level of difficulty when attempting to crack a steganography algorithm when the method used to encode is not initially given to you. Because of this, many institutions use steganography due to its robustness and capacity of data that can be hidden \cite{seethalakshmi_security_2016}. Steganography can also be used maliciously, such as hiding malware within a link that releases a virus when the user clicks on it.  

In order to decrypt the hidden images within the cover images in this project, a CycleGAN is used, which is a common architecture used in image generation \cite{welander_generative_2018}. The CycleGAN in this project takes an encoded image (one image hidden in another image) and deciphers what the hidden image is within the cover image. The images in this project were encrypted by taking the least significant RGB bits in the cover image and swapping them with the most significant RGB bits in the hidden image. Various RGB bit values (7-3 bits) were used when encrypting the images known as the LSB algorithm. Pix2pix is also used within the CycleGAN, which adds a U-Net structure to the CycleGAN as well as adding a loss function that quantifies the difference between a real and generated image. This allows the generator to more efficiently generate images similar to the input images \cite{meng_steganography_2019}. An Autoencoder, which is an unsupervised neural network that is used to reconstruct input data, is also used in this project in order to measure its effectiveness against the original architecture, which is a cycleGAN. In order to gain maximum accuracy, we utilized Bayesian Optimization, which is a methodology that globally optimizes functions with costly evaluations. This will allow the model to produce more accurate decoded images. 

Our learning technique of various bit sizes allowed for a single CycleGAN model to understand how to work with all bit sizes used with the LSB algorithm. This allows for a single model rather than 9 independent models to solve the LSB algorithm.

The method used within this paper to decode the hidden images is successful. The model successfully decodes the hidden images when passed in a training set of images encoded with a random number of RGB bit values ranging from seven to three. After training, it successfully decodes images that were encoded with a random number of RGB bit values within the same range. 

\section{Conclusion}
Our CycleGAN model was able to learn how to crack the LSB steganography algorithm with significant success as seen by the figures above. This means that our algorithm can be employed to protect people against harmful agents that use the internet and steganography algorithms to take advantage of naive users. Although the success was not perfect, there is clear indication that further work could yield better results. There are several avenues of research that can be explored to further improve steganography cracking. 

There are other architectures that could be employed which could potentially increase the accuracy of this model when decoding the images, such as Long Short-Term Memory. Different architectures provide different meaningful advantages, and exploring them could be fruitful. The bit size variation training technique can be taken further and combined with Bayesian Optimization to provide a more powerful model. Different training techniques can be employed. It is clear that the model can be expanded in various ways.

Our work can be employed to protect people against malicious uses of steganography or can be expanded more with the numerous avenues of available research.

\bibliographystyle{unsrt}  
\bibliography{khan}
\end{document}